\documentclass[a4paper, 11pt, oneside, onecolumn, openany]{article}

\usepackage[english]{babel} 
\usepackage{microtype} 
\usepackage{xspace} 
\usepackage[utf8]{inputenc}	
\usepackage[T1]{fontenc} 
\usepackage{float}

\usepackage{amsfonts,amsmath,amssymb,bm,mathtools,slashed} 

\usepackage{multicol} 
\usepackage[dvipsnames,table]{xcolor}	

\definecolor{indigo}{rgb}{0.0, 0.25, 0.42}
\newcommand{\myColor}[0]{indigo}

\colorlet{darkBlue}{blue!45!black}
\colorlet{linkColor}{blue!80!black}
\usepackage{hyperref} 
\hypersetup{
	colorlinks, 
	bookmarksopen, 
	bookmarksnumbered,
	pdftoolbar=false, 
	pdfmenubar=false, 
	pdffitwindow=false, 
	pdfstartview={FitH},	 
	citecolor=\myColor, 
	linkcolor=\myColor,	
	urlcolor=\myColor, 
	pdftitle={HighPT}, 
	pdfsubject={High-pT tail observables}, 
	pdfauthor={Felix Wilsch}, 
	pdfnewwindow=true, 
	linktoc=all
}

\usepackage{geometry}
\usepackage{fancyhdr}

\fancypagestyle{main}{%
    \fancyhf{}
    \setcounter{page}{1}
    \pagenumbering{arabic}
	\cfoot{--~\thepage~--}
	
}

\fancypagestyle{intro}{%
    \fancyhf{}
    \setcounter{page}{1}
    \pagenumbering{roman}
	\cfoot{--~\thepage~--}
	
}

\usepackage{titlesec}
\titleformat{\section}{\color{\myColor}\Large\normalfont\bfseries}{\thesection}{1.0em}{}
\titleformat{\subsection}{\color{\myColor}\large\normalfont\bfseries}{\thesubsection}{1.0em}{}
\titleformat{\subsubsection}{\color{\myColor}\normalfont\bfseries}{\thesubsubsection}{1.0em}{}

\makeatletter
\g@addto@macro\bfseries{\boldmath} 
\makeatother

\usepackage[sort&compress,numbers,merge]{natbib}
\addto\captionsenglish{}

\usepackage{graphicx}
\usepackage{epstopdf}
\graphicspath{{./Figures/}{./figures/}}
\usepackage{caption,subcaption}
\DeclareCaptionFormat{custom}
{%
    \textbf{#1#2}\textit{\sl\small #3}
}
\usepackage{booktabs} 
\usepackage{longtable} 
\usepackage{multirow} 
\renewcommand{\arraystretch}{1.2}

\usepackage[shortlabels]{enumitem}
\setlist{itemsep=.1em,topsep=.5em}
\SetEnumerateShortLabel{i}{\textit{\roman*}}

\usepackage[normalem]{ulem}
\usepackage{fontawesome}
\usepackage{bbold}
\usepackage{lipsum}

\usepackage{mmacells}
\mmaSet{
  morefv={gobble=2},
  linklocaluri=mma/symbol/definition:#1,
  morecellgraphics={yoffset=1.9ex},
  moredefined={
  EventYield, CrossSection, DifferentialCrossSection, 
  ChiSquareLHC, 
  MatchToSMEFT, SubstituteFF, SelectTerms, LHCSearch, ParallelHighPT, GetParameters, DefineParameters, PythonExport, WCxf, InitializeModel, HighPTLogo, DefineBasisAlignment,
  e, \(\nu\), u, d, WC, Coupling, FF, Mass, CKM, Vckm, Width, Param, Coefficients, EFTorder, OperatorDimension, OutputFormat, PTcuts, MLLcuts, Luminosity, SM, CombineBins, RescaleError, EFTscale, Mediators,
  Scalar, Tensor, Vector, DipoleL, DipoleQ
  }
}

\usepackage{tikz}

\newcommand{\HighPT}[0]{{\texttt{HighPT}}\xspace}
\newcommand{\Mathematica}[0]{{\texttt{Mathematica}}\xspace}
\definecolor{DarkGray}{gray}{0.40}

\usepackage{lmodern}

\numberwithin{equation}{section} 
\allowdisplaybreaks 
\numberwithin{figure}{section} 
\numberwithin{table}{section} 

\newcommand{\cmd}[1]{{\texttt{#1}}\xspace}
\newcommand{\dd}{\mathop{}\!\mathrm{d}}
\newcommand{\braces}[1]{\left\lbrace #1 \right\rbrace}
\newcommand{\brackets}[1]{\left( #1 \right)}
\newcommand{\squarebrackets}[1]{\left[ #1 \right]} 

\newcommand{\abs}[1]{\left\lvert #1 \right\rvert}

\newcommand{\cF}[0]{\mathcal{F}}

\newcommand{\cO}[0]{\mathcal{O}}
\newcommand{\cL}{{\mathcal L}}
\newcommand{\cC}{{\mathcal C}}
\newcommand{\cS}{{\mathcal S}}
\newcommand{\cT}{{\mathcal T}}
\newcommand{\cU}{{\mathcal U}}

\newcommand{\SMrep}[3]{({\bf #1},\,{\bf #2},\,#3)}
\newcommand{\SMrepbar}[3]{({\bf \bar#1},\,{\bf #2},\,#3)}

\newcommand\blfootnote[1]{%
  \begingroup
  \renewcommand\thefootnote{}\footnote{#1}%
  \addtocounter{footnote}{-1}%
  \endgroup
}

\begin{document}

\thispagestyle{empty}

\renewcommand*{\thefootnote}{\fnsymbol{footnote}} 

\begin{center} 
\begin{minipage}{15.5cm}
\vspace{-0.7cm}
\begin{flushright}
{\footnotesize \texttt{
ZU-TH-29/22
}}
\end{flushright}
\vspace{10mm}
\end{minipage}
\end{center}

\vspace{-2.1cm}
\includegraphics[width=0.35\textwidth]{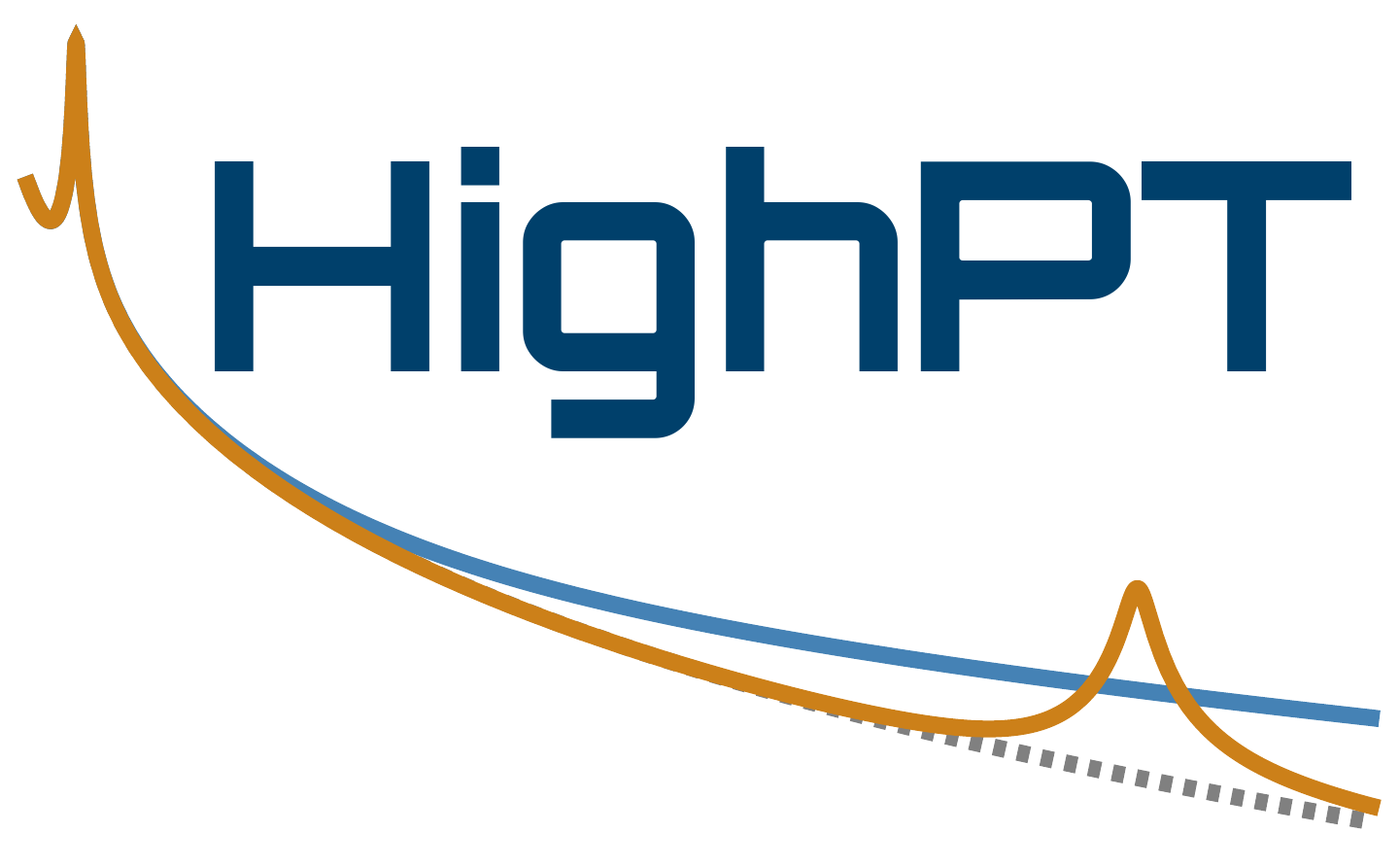}
\vspace{+0.50cm}

\begin{center}
{
	\bfseries \LARGE
    \HighPT: \\[0.25cm]
    \Large
    \resizebox{\textwidth}{!}{A Tool for high-$p_T$ Drell-Yan Tails Beyond the Standard Model}
}
\\[0.35cm]
\textcolor{\myColor}{\rule{\textwidth}{2.2pt}}
\\[0.4cm]
{ 
	\scshape 
	{L.~Allwicher${}^{\,\bullet}$,~D.~A.~Faroughy${}^{\,\bullet}$, F.~Jaffredo${}^{\,\bullet\bullet}$}, 
    {O.~Sumensari${}^{\,\bullet\bullet}$, F.~Wilsch${}^{\,\bullet}$\footnote{\href{mailto:felix.wilsch@physik.uzh.ch}{felix.wilsch@physik.uzh.ch}}}
}
\\

\vspace{0.7cm}

{
	\small
	\em
	$^\bullet$Physik-Institut, Universit\"at Z\"urich, CH-8057 Z\"urich, Switzerland
	\\
    ${}^{\bullet\bullet}$IJCLab, P\^ole Th\'eorie (Bat. 210), CNRS/IN2P3 et Universit\'e, Paris-Saclay, 91405 Orsay, France
}
\\[.4cm]
{
	\it \today
}
\\[0.5cm]
\end{center}

\setcounter{footnote}{0}
\renewcommand*{\thefootnote}{\arabic{footnote}}%
\suppressfloats	

\vspace{0.5cm}
\begin{abstract}
\vspace{+.01\textheight}
\noindent
\HighPT is a \Mathematica package for the analysis of high-energy data of semileptonic transitions at hadron colliders. 
It allows to compute high-$p_T$ tail observables for semileptonic processes, i.e.~Drell-Yan cross sections, for dilepton and monolepton final states at the LHC. 
These observables can be calculated at tree level within the Standard Model Effective Field Theory, including the relevant operators up to dimension eight to ensure a consistent description of the cross section including terms of $\cO(\Lambda^{-4})$ in the cutoff scale~$\Lambda$.
For New Physics models with new mediators that can be resolved at LHC energies, \HighPT can also account for the full propagation effects of these new bosonic states at tree level.
Using the available data from the high-$p_T$ tails in the relevant LHC run-II searches by the ATLAS and CMS collaborations,
\HighPT can also construct the corresponding likelihoods for all possible flavors of the leptonic final states.
As an illustration, we derive and compare constraints on Wilson coefficients at different orders in the Effective Field Theory expansion, and we investigate lepton flavor violation for the {$S_3$}~leptoquark model.
The \HighPT code is publicly available at~\href{https://github.com/HighPT/HighPT}{\faicon{github}}.
\blfootnote{\href{https://github.com/HighPT/HighPT}{\faicon{github} https://github.com/HighPT/HighPT}}
\end{abstract}


\newpage
\pagestyle{intro}

\newpage
{
    \hypersetup{linkcolor=black}
	\addtocontents{toc}{\protect\hypertarget{toc}{}} 
	\tableofcontents 
}


\newpage
\pagestyle{main}

\section{Introduction}

Drell-Yan production is a powerful probe for New Physics~(NP) effects, both in resonant and non-resonant searches at the LHC. 
Under the assumption of a mass gap between the electroweak scale and the NP states, these new effects can be parametrized by means of an Effective Field Theory~(EFT) approach, using the so-called Standard Model EFT~(SMEFT)~\cite{Buchmuller:1985jz,Grzadkowski:2010es}. In this case, the Drell-Yan cross section can receive energy-enhanced contributions, which can be observed as deviations from the Standard Model~(SM) expectation in the tails of momentum-dependent~distributions. Similarly, a new light degree of freedom could also manifest itself in the tails as a resonance ($s$-channel), or modify the shapes of the distributions ($t$- and $u$-channel).
The \Mathematica package \HighPT is designed to provide a complete framework to study these effects.

The measurement of Drell-Yan tails at the LHC has also been demonstrated to be an effective probe of flavored transitions~\cite{Allwicher:2022}. The energy enhancement of the $pp\to\ell\ell^\prime$ and $pp\to\ell\nu$ cross sections can overcome the suppression from Parton Distribution Functions (PDFs), allowing us to probe semileptonic effective operators with different quark and lepton flavors. These constraints are complementary to the ones derived from low-energy flavor-physics observables and they have already been proven to be useful in many cases~\cite{Cirigliano:2012ab,Chang:2014iba,Faroughy:2016osc,Greljo:2017vvb,Cirigliano:2018dyk,Greljo:2018tzh,Fuentes-Martin:2020lea,Endo:2021lhi,Marzocca:2020ueu,Jaffredo:2021ymt,Angelescu:2020uug}. While several tools exist to simplify the phenomenology of low-energy flavor observables in the SM and beyond~\cite{Straub:2018kue,Aebischer:2018iyb,EOSAuthors:2021xpv,DeBlas:2019ehy,Mahmoudi:2009zz}, a similar framework for flavor-physics observables at high-$p_T$ was still missing.

The main challenge of reinterpretations of LHC data is the Monte Carlo simulation of detector effects, which is needed to reproduce the event selection of a given experimental search. This procedure is numerically expensive, and it needs to be repeated for each specific LHC search and NP scenario. The purpose of \HighPT is to facilitate this procedure by using the relevant experimental acceptances and efficiencies that have been computed beforehand, through the recast of recent experimental searches made by the ATLAS and CMS collaborations with the full LHC run-II data, for several well-motivated NP scenarios and, most importantly, with arbitrary flavor structures. In this way, a $\chi^2$~likelihood can be constructed by comparing the expected (binned) distributions with the data that the experimental collaborations provide. The user can then use the likelihood function, which can also be exported as a {\tt python} function, using the \cmd{WCxf} format~\cite{Aebischer:2017ugx} that is also compatible with various other phenomenological codes, to fit various NP scenarios to the data.

The searches included in \HighPT are dilepton and monolepton production, for all possible flavor combinations i.e.~$\ell\ell$~\cite{ATLAS:2020zms,CMS:2021ctt} and $\ell\nu$~\cite{ATLAS:2021bjk,ATLAS:2019lsy} for $\ell=e,\mu,\tau$, as well as $\tau\mu$, $\tau e$ and $\mu e$~\cite{CMS:2022fsw}.
The NP scenarios supported by \HighPT can be summarized in three main categories, which can be selected alternatively:
\begin{itemize}
    \item \textbf{SMEFT}: All energy-enhanced operators contributing to Drell-Yan processes up to dimension eight are included, with the most general flavor indices. Thus, \HighPT can consistently compute the relevant cross sections considering contributions up to~$\cO(\Lambda^{-4})$ in the EFT power counting. 
    \item \textbf{Simplified models}: These include many different bosonic mediators, with couplings to the SM fermions and a generic flavor structure, such as scalar and vector leptoquarks, as well as colorless bosonic mediators.\,\footnote{Colorless mediators are not fully supported in version \cmd{1.0.0} but will be added in near future.}
    
    \item \textbf{Form factors}: Generic NP parametrization of the Drell-Yan amplitude based on Lorentz and $SU(3)_c\otimes U(1)_{\rm em}$ gauge invariance.
\end{itemize}

The remainder of this paper is organized as follows. In sec.~\ref{sec:cross-section} we briefly describe the details of the computation of the hadronic cross section for the relevant Drell-Yan processes. Sec.~\ref{sec:HighPT} contains the main features and functionalities of the \HighPT package, with several code examples. In sec.~\ref{sec:examples}, we demonstrate the capabilities of package by constraining SMEFT Wilson coefficients at different orders in the EFT series for a simple example, and we analyse the constraints on the $S_3$~scalar leptoquark couplings arising from searches for lepton flavor conserving and lepton flavor violating final states. We conclude and discuss our future prospects in sec.~\ref{sec:conclusion}. In the three appendices we present a list of all other variables defined in \HighPT, a discussion on the statistical treatment of LHC data, and a summary of our conventions.


\section{Framework} 
\label{sec:cross-section}

In this section we summarize the computation of the tree-level cross section for neutral and charged Drell-Yan production at hadron colliders such as the~LHC. 
For more details we refer to ref.~\cite{Allwicher:2022}, where the formalism is discussed in detail.

\subsection{Drell-Yan form factors}

We consider the $2\to2$ scattering amplitude~$\mathcal{A}$ for neutral and charged Drell-Yan production $\overline{q}_i q^{\prime}_j \to \overline{\ell}_\alpha \ell^{\prime}_\beta$, where $q^{(\prime)} \in \{u,d\}$ and $\ell^{(\prime)} \in \{\nu,e\}$, and where Greek indices~$\alpha,\beta$ run over lepton flavors whereas Latin indices~$i,j$ run over quark flavors, except for the top quark. The Drell-Yan amplitude in the fermion mass eigenbasis can be generally decomposed as follows~\cite{Allwicher:2022}:
\begin{align}
\begin{split}
[{\mathcal{A}}]_{\alpha\beta ij} 
&\equiv {\mathcal{A}} \brackets{\overline{q}_i q^{\prime}_j \to \overline{\ell}_\alpha \ell^{\prime}_\beta}
\\
&=
\frac{1}{v^2} \sum_{X,Y} \Bigg\{ \brackets{\overline{\ell}_\alpha \mathbb{P}_X \ell^{\prime}_\beta} \brackets{\overline{q}_i \mathbb{P}_Y q^{\prime}_j} \squarebrackets{\mathcal{F}_S^{XY,qq^{\prime}}\!\!(\hat{s},\hat{t})}_{\alpha\beta ij}
\\
&\qquad+\brackets{\overline{\ell}_\alpha \gamma_\mu \mathbb{P}_X \ell^{\prime}_\beta} \brackets{\overline{q}_i \gamma^\mu \mathbb{P}_Y q^{\prime}_j} \squarebrackets{\mathcal{F}_V^{XY,qq^{\prime}}\!\!(\hat{s},\hat{t})}_{\alpha\beta ij}
\\[0.2em]
&\qquad+\brackets{\overline{\ell}_\alpha \sigma_{\mu\nu} \mathbb{P}_X \ell^{\prime}_\beta} \brackets{\overline{q}_i \sigma^{\mu\nu} \mathbb{P}_Y q^{\prime}_j} \delta^{XY} \squarebrackets{\mathcal{F}_T^{XY,qq^{\prime}}\!\!(\hat{s},\hat{t})}_{\alpha\beta ij}
\\[0.2em]
&\qquad+\brackets{\overline{\ell}_\alpha \gamma_\mu \mathbb{P}_X \ell^{\prime}_\beta} \brackets{\overline{q}_i \sigma^{\mu\nu} \mathbb{P}_Y q^{\prime}_j} \frac{i k_\nu}{v} \squarebrackets{\mathcal{F}_{D_q}^{XY,qq^{\prime}}\!\!(\hat{s},\hat{t})}_{\alpha\beta ij}
\\[0.2em]
&\qquad+\brackets{\overline{\ell}_\alpha \sigma_{\mu\nu} \mathbb{P}_X \ell^{\prime}_\beta} \brackets{\overline{q}_i \gamma^\mu \mathbb{P}_Y q^{\prime}_j} \frac{i k^\nu}{v} \squarebrackets{\mathcal{F}_{D_\ell}^{XY,qq^{\prime}}\!\!(\hat{s},\hat{t})}_{\alpha\beta ij} \Bigg\} \,.
\label{eq:amplitude}
\end{split}
\end{align}
Here $\mathcal{F}_I^{XY,qq^{\prime}}\!(\hat{s},\hat{t})$ denote different form factors for each possible Lorentz structure, i.e.~\textit{scalar}, \textit{vector}, \textit{tensor}, \textit{quark dipoles}, and \textit{lepton dipole} $(I=S,V,T,D_q,D_\ell)$. The form factors depend on the partonic Mandelstam variables defined as~$\hat s= k^2 =(p_q+p_{\bar q})^2$, $\hat t=(p_{q}-p_{\bar\ell})^2$, and $\hat u=(p_{q}-p_{\ell^{\prime}})^2$, where all dependence on~$\hat{u}$ can be removed using $\hat{u} = - \hat{s} - \hat{t}$ for massless particles, which we assume throughout this work. Moreover, the form factors depend on the chiralities~$X,Y \in\{L,R\}$ of the lepton and quark bilinears, respectively, and four flavor indices. The chirality projectors are denoted~$\mathbb{P}_{L,R}$. 
Eq.~\eqref{eq:amplitude} is the most general form-factor decomposition of the amplitude assuming Lorentz invariance and invariance under the gauge group~$\mathcal{G}=SU(3)_c \otimes U(1)_\mathrm{em}$. 

Following ref.~\cite{Allwicher:2022}, we decompose each form factor into a regular and a singular piece,
\begin{align}\label{eq:FF}
\cF_{I}(\hat  s,\hat{t})\ =\ \cF_{I,\, \rm Reg}(\hat s,\hat{t})\ +\ \cF_{I,\, \rm Poles}(\hat s,\hat{t})\,,  
\end{align}
which we parametrize as follows:
\begin{align}\label{eq:regular}
    \cF_{I,\,\rm Reg}(\hat  s,\hat{t})&=\sum_{n,m=0}^\infty \cF_{I \,(n,m)}\,\bigg[\frac{\hat s}{v^2}\bigg]^n\bigg[\frac{\hat t}{v^2}\bigg]^m\,,\\
\cF_{I,\,\rm Poles}(\hat  s,\hat{t}) &= \sum_a\frac{v^2\, \cS_{\,I\,(a)} }{\hat{s}-\Omega_a}
\ +\  \sum_b\frac{v^2\, \cT_{\,I\,(b)} }{\hat{t}-\Omega_b}
\ -\  \sum_c\frac{v^2\, \cU_{\,I\,(c)} }{\hat{s} + \hat {t} +\Omega_c}\,.\label{eq:poles}
\end{align} 
Here, the analytic functions $\cF_{I,\, \rm Reg}$ encode the interactions from heavy and thus unresolved degrees of freedom, e.g. four-fermion contact interactions, while the singular functions $\cF_{I,\, \rm Poles}$ contain simple poles $\Omega_k = m_k^2 - i m_k\Gamma_k$ in each Mandelstam plane, describing resolved degrees of freedom with masses $m_{a,b,c}$ and widths $\Gamma_{a,b,c}$ that can be exchanged at tree level in the $s$-, $t$- and $u$-channels, respectively. The power expansion\,\footnote{This is equivalent to the derivative expansion for higher-dimensional effective operators.} in the Mandelstam variables in eq.~\ref{eq:regular} is valid provided $v^2,|\hat{s}|,|\hat{t}|\ll\Lambda^2$, i.e. within the validity range of the EFT approach.

Specific models contributing to Drell-Yan production at tree level can be directly matched to the form-factor parameters introduced above, i.e. the expansion coefficients $\cF_{I\,(n,m)}$ and the pole  residues~$\cS_{\,I\,(a)}$, $\cT_{\,I\,(b)}$, and $\cU_{\,I\,(c)}$. For instance, the SM will only give rise to the vector poles $\cS_{V\, (Z)}$, $\cS_{V\, (\gamma)}$ in neutral currents and $\cS_{V\, (W)}$ in charged currents, through the $\gamma$, $Z$ and $W^\pm$ gauge bosons, respectively. In the SMEFT, there will be energy-enhanced contributions from $d=6$ and $d=8$ operators that give rise to the regular form-factor coefficients $\cF_{I\,(n,m)}$ with $n,m\leq1$\,\footnote{At dimension six only coefficients with $(n,m)=(0,0)$ are generated, whereas $d=8$ operators yield contributions to $(n,m)\in\{(0,0),~(1,0),~(0,1)\}$. More generally, an operator of mass dimension~$d$ can only generate terms with $d\geq 6 + 2(n+m)$.} and additional contributions to the $s$-channel pole residues $\cS_{\,I\,(a)}$ coming from modified SM gauge boson couplings~($I=V$) and from dipole operators~($I=D_{\ell,q}$). The pole coefficients $\cT_{\,I\,(b)}$ and $\cU_{\,I\,(c)}$ are absent for the SMEFT, but can appear for concrete models, e.g.~leptoquarks can contribute to $\cT_{\,S,V,T}$ and $\cU_{\,S,V,T}$, whereas colorless mediators can contribute to $\cS_{\,S,V}$.

The form factors in eq.~\eqref{eq:amplitude} are expressed in the fermion mass basis.
To match these form factors onto SMEFT Wilson coefficients or couplings of ultraviolet mediators, it is convenient to rotate the form factors to the basis of weak interactions.
The relation of the fields in the mass basis to the fermions in the weak basis is given by the left-handed rotation matrices~$V_u$ and~$V_d$~as
\begin{align}
    u_L^\mathrm{mass} &= V_u \cdot u_L^\mathrm{weak} \,, 
    &
    d_L^\mathrm{mass} &= V_d \cdot d_L^\mathrm{weak} \,.
    \label{eq:qL_rotation_matrices}
\end{align}
The Cabibbo–Kobayashi–Maskawa~(CKM) matrix is then defined by $V_\mathrm{CKM}=V_u^\dagger \cdot V_d$ and we find the following rotations for the form factors
\begin{align}
    \mathcal{F}_{V,\,\mathrm{mass}}^{XL,\,qq^\prime} &\longrightarrow 
    V_{q} \cdot \cF_{V,\,\mathrm{weak}}^{XL,\,qq^\prime} \cdot V_{q^\prime}^\dagger \,, 
    \label{eq:FF_V_rotation}
    \\[0.2em]
    \mathcal{F}_{I \neq V,\,\mathrm{mass}}^{XL,\,qq^\prime} &\longrightarrow
    \cF_{I \neq V,\,\mathrm{weak}}^{XL,\,qq^\prime} \cdot V_{q^\prime}^\dagger \,,
    \label{eq:FF_L_rotation}
    \\[0.2em]
    \mathcal{F}_{I \neq V,\,\mathrm{mass}}^{XR,\,qq^\prime} &\longrightarrow
    V_q \cdot \cF_{I \neq V,\,\mathrm{weak}}^{XR,\,qq^\prime} \,.
    \label{eq:FF_R_rotation}
\end{align}
The Pontecorvo–Maki–Nakagawa–Sakata~(PMNS) matrix could be incorporated in this framework analogously to eqs.~\eqref{eq:qL_rotation_matrices}--\eqref{eq:FF_R_rotation}. However, since we are not including right-handed neutrinos at the moment, the PMNS can be set to the unit matrix.
In the remainder of this work we will consider all flavor indices to correspond to weak eigenstates, unless stated otherwise.

\subsection{Partonic cross section}
The partonic cross section~$\hat{\sigma}(\hat{s})$ for the process $\overline{q}_i q^{\prime}_j \to \overline{\ell}_\alpha \ell^{\prime}_\beta$ can be expressed as an integration over the Mandelstam variable~$\hat{t}$,
\begin{align}
\label{eq:partonic-xsection}
\hat\sigma (\overline{q}_i q^{\prime}_j \to \overline{\ell}_\alpha \ell^{\prime}_\beta;\,\hat s)
&= \frac{1}{16\pi} \frac{1}{\hat{s}^2} \int_{\hat{t}_{-}}^{\hat{t}_{+}} \dd\hat{t} \, \overline{\abs{\squarebrackets{\mathcal{A}}_{\alpha\beta ij}}^2}
\\
&= \frac{1}{48\pi v^4} \int_{\hat{t}_{-}}^{\hat{t}_{+}} \dd\hat{t} \, \sum_{X,Y} \sum_{I,J} M_{IJ}^{XY}\!\!\brackets{\hat{s},\hat{t}} \squarebrackets{\cF_I^{XY}\!\!\brackets{\hat{s},\hat{t}}}_{\alpha\beta ij} \squarebrackets{\cF_J^{XY}\!\!\brackets{\hat{s},\hat{t}}}_{\alpha\beta ij}^\ast \,,\nonumber
\end{align}
where the bar over the squared amplitude denotes the averaged sum over spin and color. The dimensionless ${5 \times 5}$~matrices $M^{XY}$ are given by
\begin{align}
M^{XY}(\hat s,\hat t)=\begin{pmatrix}
M_{VV}^{XY}(\hat t/\hat s) & 0 &  0 & 0 & 0 \\
0 & M_{SS}^{XY} & M_{ST}^{XY}(\hat t/\hat s) & 0 & 0 \\
0 & M_{ST}^{XY}(\hat t/\hat s) & M_{TT}^{XY}(\hat t/\hat s)  & 0 & 0 \\
0 & 0 & 0 & \frac{\hat s}{v^2} M_{DD}^{XY}(\hat t/\hat s)& 0 \\
0 & 0 & 0 & 0 & \frac{\hat s}{v^2}M_{DD}^{XY}(\hat t/\hat s)
\end{pmatrix} \,,
\end{align} 
where the different matrix entries~$M_{IJ}^{XY}(\hat t/\hat s)$ are polynomials defined by
\begin{subequations}
\begin{align}
    M_{VV}^{XY}(\omega) &= (1+2\omega)\delta^{XY}+\omega^2\,, \\
    M_{SS}^{XY}  &= 1/4\,, \\
    M_{ST}^{XY}(\omega) &= -(1+2\omega)\delta^{XY} \,, \\
    M_{TT}^{XY}(\omega) &= 4(1+2\omega)^2\delta^{XY}\,, \\
    M_{DD}^{XY}(\omega) &= -\omega(1+\omega)\,,
\end{align}
\end{subequations}
which only depend on the emission angle~$\theta^\ast$ of the lepton with respect to the initial state quarks, since we have~$\smash{\omega=\hat{t}/\hat{s}=\frac{1}{2}(\cos\theta^\ast-1)}$.

\paragraph{Angular integration}
Integrating over the full emission angle~$\cos \theta^\ast = 1 + 2\, \hat{t}/\hat{s}$ is equivalent to integrating over $\hat{t}$ and fixing the integration boundaries in eq.~\eqref{eq:partonic-xsection} as~$\hat{t}_{+}=0$ and~$\hat{t}_{-}=-\hat{s}$. In the case where the angular integration is constrained by cuts on the transverse momentum~$p_T$ of the emitted leptons, i.e.~by requiring $p_{T_\mathrm{min}} \leq p_T \leq p_{T_\mathrm{max}}$, the $\hat{t}$~integral in eq.~\eqref{eq:partonic-xsection} must be replaced by
\begin{align}\label{eq:t-integral_shift}
    \int_{\hat{t}_{-}}^{\hat{t}_{+}} \dd\hat{t} \longrightarrow \int_{\hat{t}_1^+}^{\hat{t}_2^+} \dd\hat{t} + \int^{\hat{t}_1^-}_{\hat{t}_2^-} \dd\hat{t} \,,
\end{align}
with the new integration boundaries given by\,\footnote{Notice that for $\hat{s} < 4 p^2_{T_\mathrm{max}}$ we find $\hat{t}_2^- = \hat{t}_2^+$, whereas for $\hat{s}<4 p^2_{T_\mathrm{min}}$ the cross section vanishes. Taking the limit $p_{T_\mathrm{min}}\to 0$ and $p_{T_\mathrm{max}}\to\infty$ yields again the integration boundaries for the full angular integration.}
\begin{align}
\hat{t}_1^\pm (\hat{s}) &= - \frac{\hat{s}}{2} \brackets{1 \pm \sqrt{1- \min \braces{1,\ 4\frac{p^2_{T_\mathrm{min}}}{\hat{s}}}}} \,, & 
\hat{t}_2^\pm (\hat{s}) &= - \frac{\hat{s}}{2} \brackets{1 \pm \sqrt{1- \min \braces{1,\ 4\frac{p^2_{T_\mathrm{max}}}{\hat{s}}}}} \,.
\label{eq:t-cuts}
\end{align}
Notice that the scalar-tensor interference term~$M_{ST}=-(1+2\hat{t}/\hat{s})=-\cos\theta^\ast$ vanishes when integrating over any arbitrary $p_T$~bin. This more generally holds for any observable that is symmetric in the angle~$\theta^\ast-\pi/2$. To probe $M_{ST}$ one should examine, e.g.,~forward-backward asymmetries.

The $\hat{t}$~integrals can further be simplified using a partial fraction decomposition
\begin{align}
    \frac{1}{\hat{t}-\Omega_A} \, \frac{1}{\hat{t}-\Omega_B} = \frac{1}{\Omega_A - \Omega_B} \left( \frac{1}{\hat{t}-\Omega_A} - \frac{1}{\hat{t}-\Omega_B} \right)\,,
    \label{eq:partial-fraction}
\end{align}
if $\Omega_A \neq \Omega_B$, and identities such as 
\begin{align}
    \frac{\hat{t}}{\hat{t}-\Omega} = 1 + \frac{\Omega}{\hat{t}-\Omega} \,,
    \label{eq:integral-reduction}
\end{align}
which are used in \HighPT to simplify results. Similar identities also apply for $u$-channel propagators after replacing $\hat{u}=-\hat{t}-\hat{s}$. Therefore, the whole $\hat{t}$~integration can be performed fully analytically.

\subsection{Hadronic cross section}

The hadronic cross section, for a specific bin $\hat s \in [m_{\ell\ell_{\text{min}}}^2,m_{\ell\ell_{\text{max}}}^2]$ and integrated over the full $\hat{t}$ range, is given by
\begin{align}\label{eq:master-formula}
\begin{split}
\sigma\brackets{pp \to \bar{\ell}_\alpha \ell_\beta^{(\prime)};\,s}
&=
\frac{1}{48\pi v^4} \sum_{ij} \int_{m_{\ell\ell_\mathrm{min}}^2}^{m_{\ell\ell_\mathrm{max}}^2} \frac{\dd \hat{s}}{\hat{s}} L_{ij}\brackets{\frac{\hat{s}}{s}} 
 \int_{-\hat s}^{0} \dd\hat{t}
\\[0.1cm]
&\quad \times\sum_{X,Y} \sum_{I,J} M_{IJ}^{X,Y}\!\!\brackets{\hat{s},\hat{t}} \squarebrackets{\cF_I^{X,Y}\!\!\brackets{\hat{s},\hat{t}}}_{\alpha\beta ij}^\ast \squarebrackets{\cF_J^{X,Y}\!\!\brackets{\hat{s},\hat{t}}}_{\alpha\beta ij} \,,
\end{split}
\end{align}
where $L_{ij}(\tau)$ ($\tau = \hat s/s$) are the parton-parton luminosity functions defined in terms of the PDFs as
\begin{align}\label{eq:parton-luminosities}
	L_{ij} (\tau) = \tau \int_\tau^1 \frac{\dd x}{x} \squarebrackets{f_{\bar{q}_i}\brackets{x,\mu} f_{q_j}\brackets{\frac{\tau}{x},\mu} + (\bar{q}_i \leftrightarrow q_j)} \, .
\end{align}
To obtain the hadronic cross section with $p_T$~cuts, one simply needs to replace the integration limits of the $\hat t$ integral as described in eqs.~\eqref{eq:t-integral_shift} and~\eqref{eq:t-cuts}.\,\footnote{Note, also that although the $\hat{s}$~integration is performed numerically due to the presence of the parton-parton luminosity functions, the evaluation of the integrals can still be simplified by applying identities similar to eq.~\eqref{eq:partial-fraction} and~\eqref{eq:integral-reduction}.}

\subsection{High-$p_T$ tail observables}
\label{subsec:efficiencies}

\begin{figure}[t!]
    \centering
    \includegraphics[width=0.3\textwidth]{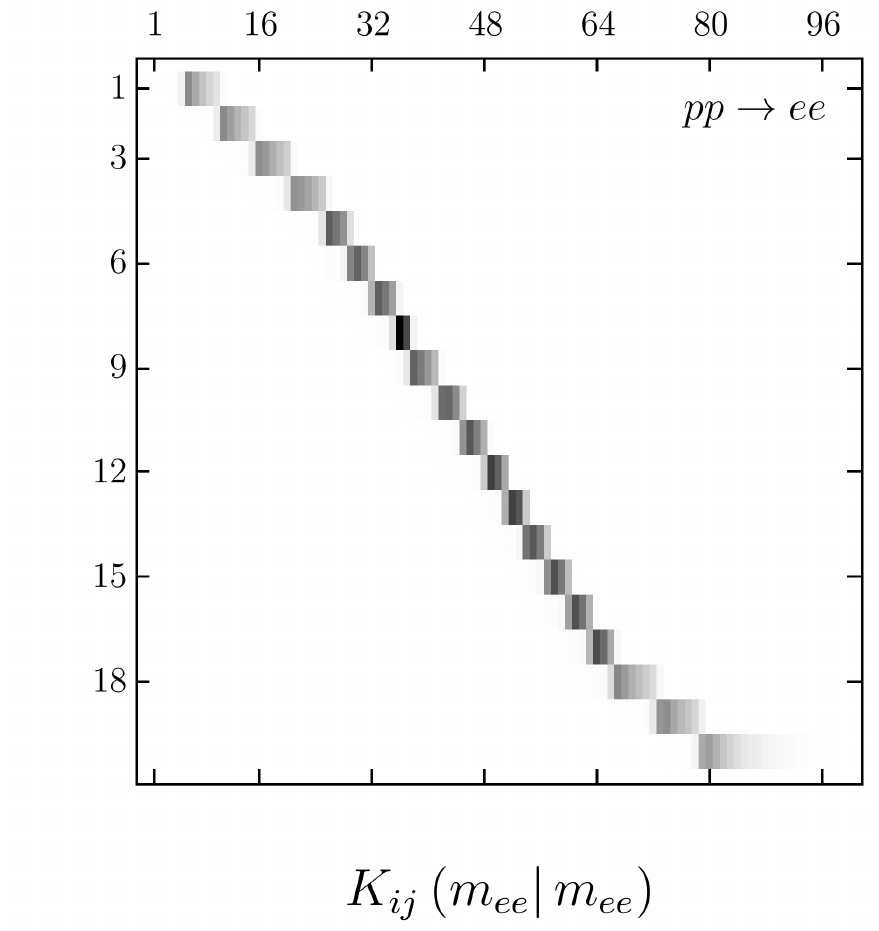}
    \includegraphics[width=0.3\textwidth]{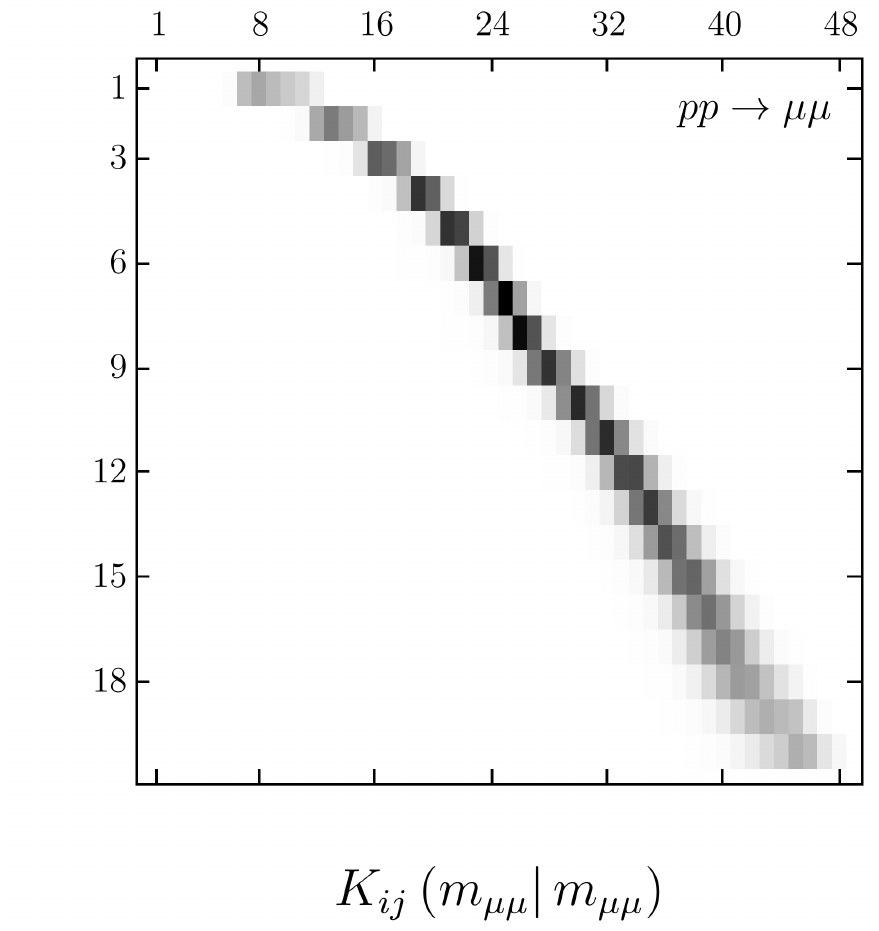}
    \includegraphics[width=0.3\textwidth]{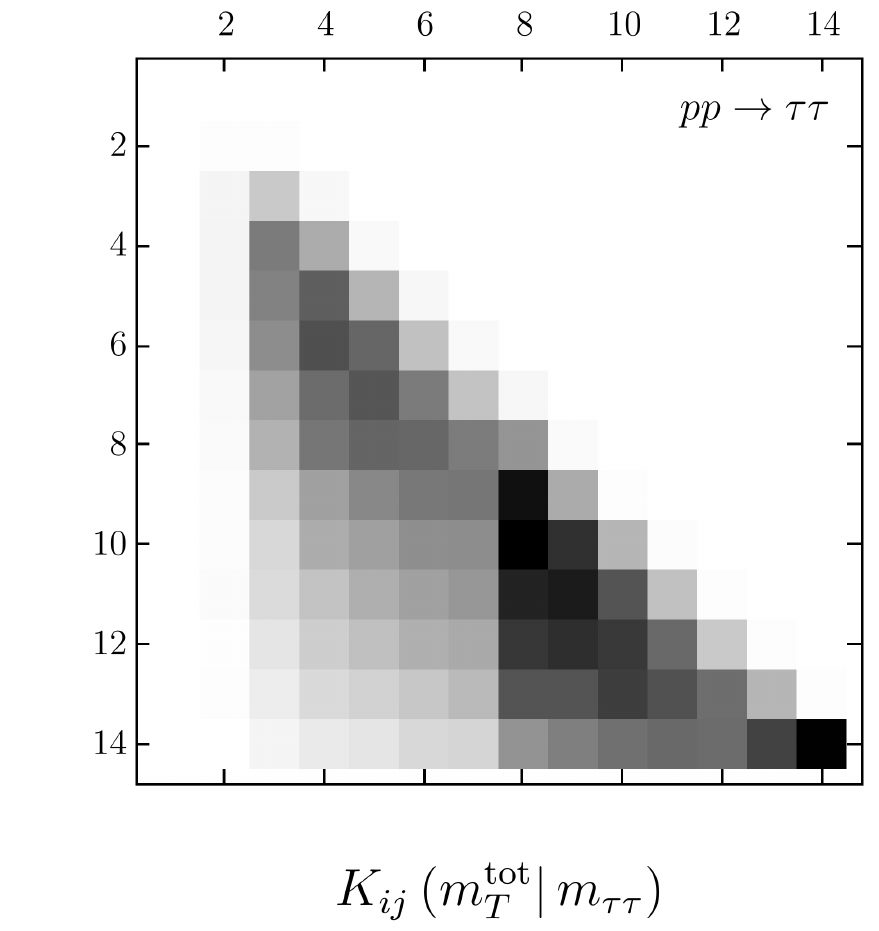}\\
    \vspace{0.5cm}
    \includegraphics[width=0.3\textwidth]{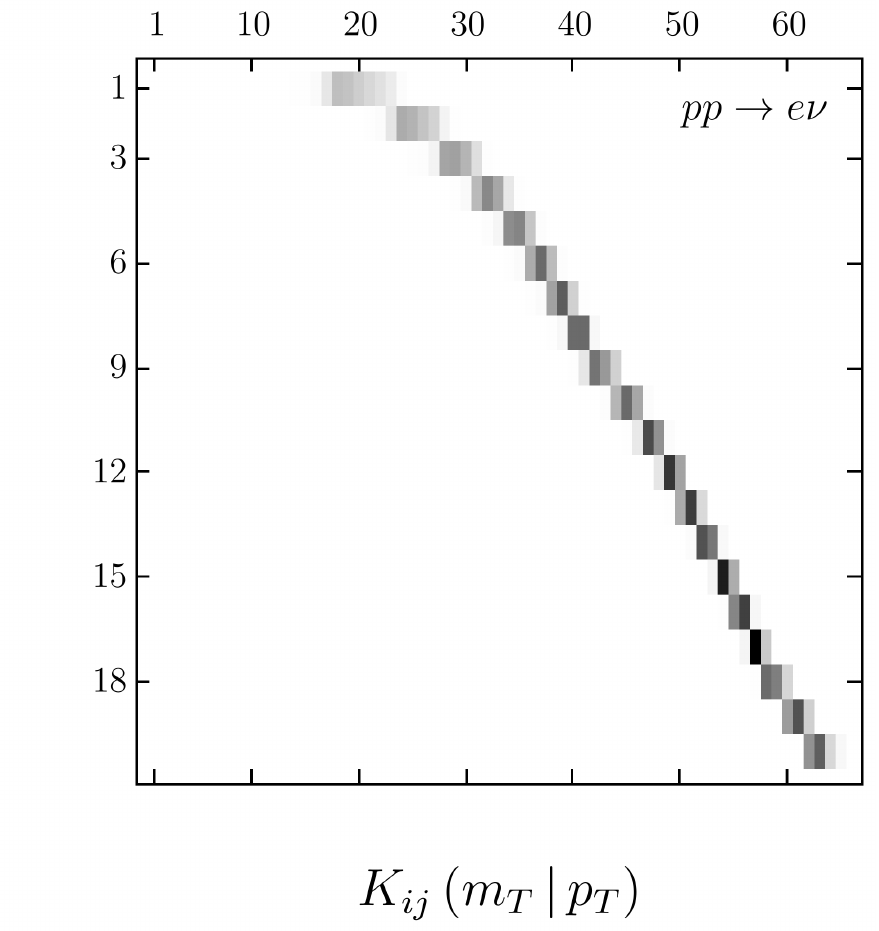}
    \includegraphics[width=0.3\textwidth]{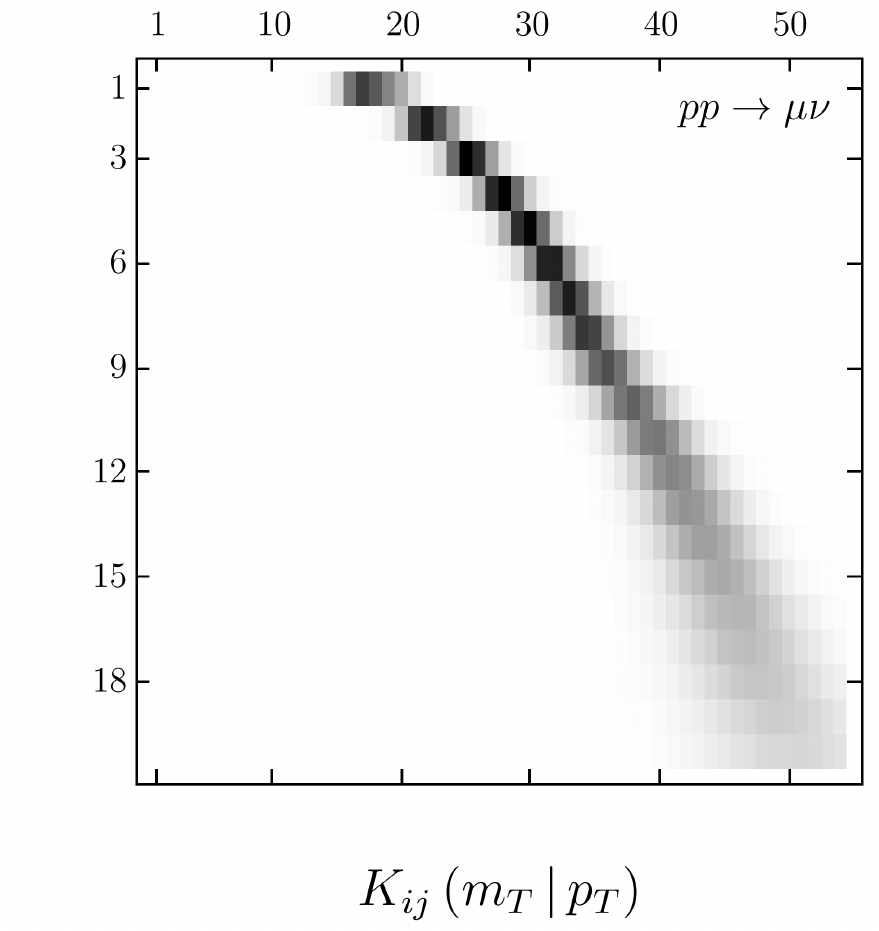}
    \includegraphics[width=0.3\textwidth]{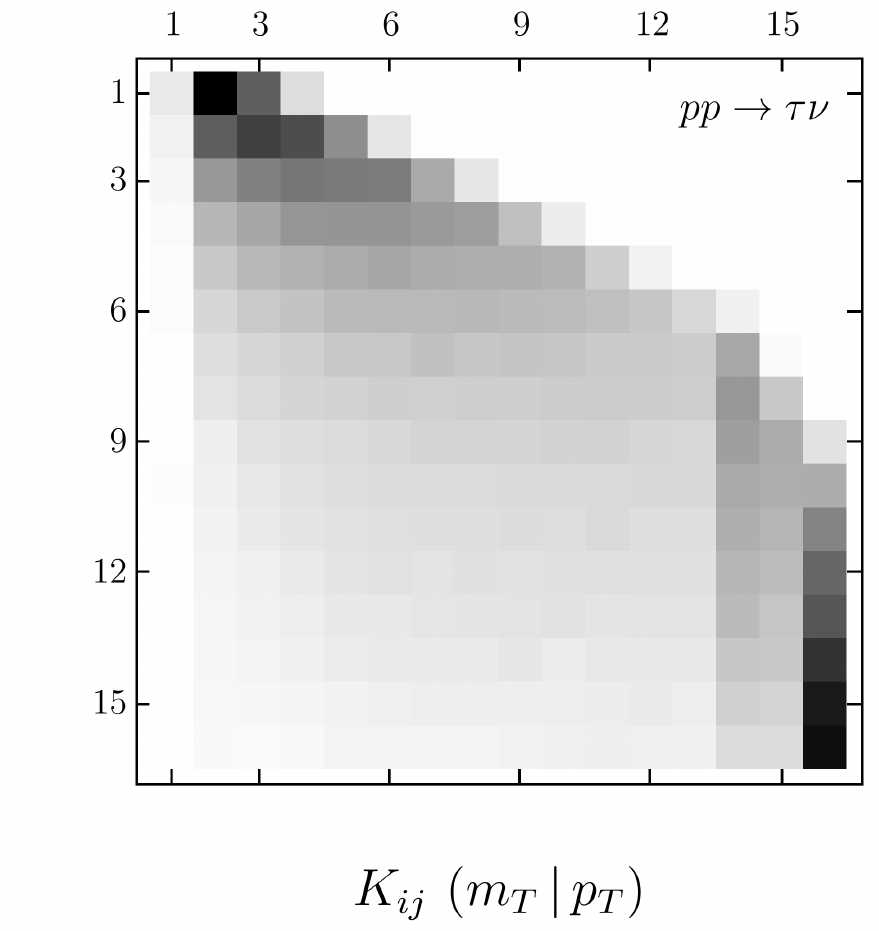}
    \caption{Kernel matrices $K_{ij}\,(x_{\rm obs}\,|\,x)$ for the monolepton and dilepton tails at the LHC listed in tab.~\ref{tab:searches}. These results are for SM contribution via off-shell $Z$- and $W$-boson for initial valence quarks.}
    \label{fig:Kernel_Matrices}
\end{figure}

The particle-level cross section for specific invariant mass~$m_{\ell\ell}$ or transverse momentum~$p_T$ bins as given in eq.~\eqref{eq:master-formula} is not directly measurable by experiments at the LHC. This is due to experimental measurements relying on particle reconstruction from objects such as isolated light leptons, tagged jets, and missing transverse momentum~$\slashed{E}_T$ for which the full information is not available experimentally due to the nature of the process or limitations of the detector.

In general, for binned distributions the particle-level observables~$x$ (with $M$~bins) can be related to the experiment-level observables~$x_\mathrm{obs}$ (with $N$~bins) by
\begin{align}
    \sigma_j(x_{\rm obs}) = \sum_{i=1}^M \, K_{ij}\,\sigma_i(x) \,,
    \label{eq:efficiency_convolution}
\end{align}
where the cross sections for both types of observable are restricted to the $i^\mathrm{th}$ and $j^\mathrm{th}$ bin, respectively.
The $N \times M$~response matrix~$K$ relating the cross section to the observables cannot be computed from first principles and has to be derived using Monte Carlo techniques. Notice that in our framework each independent combination of form factors has their own response matrix, as the experimental efficiencies depend on the kinematics and the flavor content, which differ for the different terms in the amplitude. For further details see the discussion in ref.~\cite{Allwicher:2022}. In fig.~\ref{fig:Kernel_Matrices} we display the shapes of the response matrices for the SM contribution to dilepton and monolepton tails from valence quarks. As expected, we find that for the hadronic $\tau$ searches these matrices are highly non-diagonal, mainly due to the inefficiency in reconstructing the final states due to missing energy from the neutrinos. On the other hand, for searches with final state muons and electrons the response matrices are approximately diagonal, where the off-diagonal entries arise due to energy smearing and detector effects. We find very similar results for sea-quarks.

\section{\tt HighPT}
\label{sec:HighPT}
We introduce {\HighPT} (High-$p_T$ Tails, pronounced ``hyped'') a \Mathematica package for the computation of high-$p_T$ tail observables at the LHC. 
The main objective of \HighPT is to compute the tree-level cross section for neutral- and charged-current Drell-Yan production:
\begin{align}
pp &\to \ell_\alpha^- \ell_\beta^+ \,,
&
pp &\to \ell_\alpha^- \bar{\nu}_\beta \,,
&
pp &\to \nu_\alpha \ell_\beta^+ \,.
\end{align}
These computations can be performed at tree level in the SMEFT, incorporating the relevant effective operators up to dimension eight, for a consistent truncation of the cross section, keeping terms up to $\cO(\Lambda^{-4})$, or in explicit NP models with additional heavy bosonic mediators contributing at tree level to the $s$-,~$t$- and $u$-channels.
Moreover, \HighPT allows for complex valued Wilson coefficients and coupling constants.

Computing the relevant cross sections and using recasts of the latest experimental searches, \HighPT can derive a prediction of the event yield for each bin in the respective experimental searches as described in sec.~\ref{subsec:efficiencies}. This computation can be performed both in the SMEFT, as a function of the Wilson coefficients, or in any of the NP models listed in tab.~\ref{tab:mediators}, as a function of the respective couplings to fermions.
The experimental searches available in \HighPT as of version~\cmd{1.0.0} are shown in tab.~\ref{tab:searches}, which are all based on the full run-II data gathered at LHC by the ATLAS and CMS collaborations for the relevant monolepton and dilepton final states. Specific details concerning the selection cuts and the inputs used in these experimental analyses are available in the respective ATLAS and CMS papers listed in tab.~\ref{tab:searches}.
We are planing on adding further searches such as~\cite{CMS:2022yjm,CMS:2022rbd,ATLAS:2019erb} in near future, to eventually have the full run-II data by both the CMS and ATLAS collaboration available.

\begin{table}[t!]
    \centering
    {
    \renewcommand{\arraystretch}{1.5}
    \small
    \begin{tabular}{l | c c c c c c}
        Process & \HighPT label & Experiment & Lumi. & $x_\mathrm{obs}$ & $x$ & Ref.
        \\\hline\hline
        $pp \to \tau^+\tau^-$ & \mmaInlineCell[]{Input}{"di-tau-ATLAS"} & ATLAS & $139\,\mathrm{fb}^{-1}$ & $m_T^{\rm tot}(\tau_h^1,\tau_h^2,\slashed{E}_T)$ & $m_{\tau\tau}$ & \cite{ATLAS:2020zms}
        \\
        $pp \to \mu^+\mu^-$ & \mmaInlineCell[]{Input}{"di-muon-CMS"} & CMS & $140\,\mathrm{fb}^{-1}$ & $m_{\mu\mu}$ & $m_{\mu\mu}$ & \cite{CMS:2021ctt}
        \\
        $pp \to e^+ e^-$ & \mmaInlineCell[]{Input}{"di-electron-CMS"} & CMS & $137\,\mathrm{fb}^{-1}$ & $m_{ee}$ & $m_{ee}$ & \cite{CMS:2021ctt}
        \\\hline
        $pp \to \tau^\pm\nu$ & \mmaInlineCell[]{Input}{"mono-tau-ATLAS"} & ATLAS & $139\,\mathrm{fb}^{-1}$ & $m_T(\tau_h,\slashed{E}_T)$ & $p_T(\tau)$ & \cite{ATLAS:2021bjk}
        \\
        $pp \to \mu^\pm\nu$ & \mmaInlineCell[]{Input}{"mono-muon-ATLAS"} & ATLAS & $139\,\mathrm{fb}^{-1}$ & $m_T(\mu,\slashed{E}_T)$ & $p_T(\mu)$ & \cite{ATLAS:2019lsy}
        \\
        $pp \to e^\pm\nu$ & \mmaInlineCell[]{Input}{"mono-electron-ATLAS"} & ATLAS & $139\,\mathrm{fb}^{-1}$ & $m_T(e,\slashed{E}_T)$ & $p_T(e)$ & \cite{ATLAS:2019lsy}
        \\\hline
        $pp \to \tau^\pm \mu^\mp$ & \mmaInlineCell[]{Input}{"muon-tau-CMS"} & CMS & $138\,\mathrm{fb}^{-1}$ & $m_{\tau_h\mu}^{\rm col}$ & $m_{\tau\mu}$ & \cite{CMS:2022fsw}
        \\
        $pp \to \tau^\pm e^\mp$ & \mmaInlineCell[]{Input}{"electron-tau-CMS"} & CMS & $138\,\mathrm{fb}^{-1}$ & $m_{\tau_he}^{\rm col}$ & $m_{\tau e}$ & \cite{CMS:2022fsw}
        \\
        $pp \to \mu^\pm  e^\mp$ & \mmaInlineCell[]{Input}{"electron-muon-CMS"} & CMS & $138\,\mathrm{fb}^{-1}$ & $m_{\mu e}$ & $m_{\mu e}$ & \cite{CMS:2022fsw}
    \end{tabular}
    }
    \caption{Experimental searches by the ATLAS and CMS collaborations that are available in \HighPT as of version \cmd{1.0.0}. The currently supported list of searches can also be obtained after loading \HighPT by the command \cmd{LHCSearch[]}. The labels given in the second column can be used in the routines \texttt{ChiSquareLHC} and \texttt{EventYield} to specify the given search.}
    \label{tab:searches}
\end{table}

To recast the most recent LHC searches listed in tab.~\ref{tab:searches}, we have implemented the $d=6$~\cite{Buchmuller:1985jz,Grzadkowski:2010es} and $d=8$~\cite{Li:2020gnx,Murphy:2020rsh}  SMEFT Lagrangians in \cmd{FeynRules}~\cite{Alloul:2013bka}. 
The resulting UFO model files were imported into \cmd{MadGraph5} \cite{Alwall:2014hca} to simulate statistically significant event samples for the relevant monolepton and dilepton channels, for all possible initial quark flavors, by using the \cmd{PDF4LHC15\_nnlo\_mc} PDF set~\cite{Butterworth:2015oua}. The simulations were performed for the relevant SMEFT operators and for all possible NP couplings to fermions.
The Monte Carlo samples were then showered and hadronized using \cmd{Pythia8}~\cite{Sjostrand:2014zea}, and the final-state object reconstruction and detector simulation were performed with \cmd{Delphes3}~\cite{deFavereau:2013fsa} by using parameters tuned to the respective experimental searches. 

The cross-section computation is performed analytically, following the discussion in sec.~\ref{sec:cross-section}, except for the integration over the partonic center-of-mass energy~$\hat{s}$ which has to be performed numerically. All results obtained by \HighPT can be expressed in terms of form factors, EFT Wilson coefficients, or NP couplings to fermions. \HighPT uses by default the average parton distribution functions provided by the \cmd{PDF4LHC15\_nnlo\_mc} PDF set~\cite{Butterworth:2015oua}.\,\footnote{The function \cmd{SetPDF} can be used to change the PDF set used for cross-section computations as explained in tab.~\ref{tab:FF_WC}.} The parton-parton luminosity functions have been computed according to eq.~\eqref{eq:parton-luminosities} with the help of the \cmd{ManeParse} package~\cite{Clark:2016jgm}.

Apart from the cross-section computation, \HighPT also allows to obtain the expected number of event in the bins of the available experimental searches. Furthermore, it can construct a $\chi^2$~likelihood that can be further analyzed using \Mathematica's built-in minimization and plotting routines.
Event yields and likelihoods can also be exported as a \cmd{python} file using the \cmd{WCxf} format~\cite{Aebischer:2017ugx},\,\footnote{The \cmd{WCxf} format only provides conventions for the $d=6$ operators in the Warsaw basis, for the $d=8$ operators we introduce a similar convention, and for the NP coupling constants we present our notation in tab.~\ref{tab:couplings}.} which is also used by other phenomenological codes.

In the following, we give the installation instructions and describe the main routines of \HighPT. A complete list of all remaining functions is given in the tables of app.~\ref{app:functions}. For further information, see the ancillary \Mathematica notebooks available at \href{https://github.com/HighPT/HighPT}{\color{\myColor}\faicon{github}} which contain code examples for all functions.

\subsection{Downloading and installing the package}
The \HighPT package is free software released under the MIT License and is publicly available in the GitHub repository:
\begin{center}
\href{https://github.com/HighPT/HighPT}{https://github.com/HighPT/HighPT}
\end{center}
The package can be installed in one of two ways:
\begin{enumerate}[i)]
\item \textit{Automatic installation}: The simplest way to download and install \HighPT is to run the following command in a \Mathematica notebook:
\begin{mmaCell}{Input}
  Import["https://github.com/HighPT/HighPT/raw/master/install.m"] 
\end{mmaCell}
This will download and install \HighPT in the \textit{Applications} folder of \Mathematica's base directory.
    
\item \textit{Manual installation}: The user can also manually download the package from the GitHub repository \href{https://github.com/HighPT/HighPT}{\color{\myColor}\faicon{github}}. 
In this case, the directory containing the \HighPT code must be placed in \Mathematica's path, which can be achieved specifying:
\begin{mmaCell}{Input}
  PrependTo[\$Path,"path/to/HighPT/directory"];
\end{mmaCell}
each time before loading the package.\,\footnote{We recommend placing the \HighPT folder in the \texttt{Applications} folder of \Mathematica's base directory to avoid specifying the path.}
\end{enumerate}
Once installed, the user can load \HighPT in any \Mathematica notebook by running:
\begin{mmaCell}{Input}
  << \mmaDef{HighPT\(\,\grave\,\)}
\end{mmaCell}
After loading the package, the SMEFT run mode is activated. By default cross sections are computed considering contributions up to and including $\mathcal{O}(\Lambda^{-4})$ by $d=6$ operators only assuming a cutoff scale of~$\Lambda=1\,\mathrm{TeV}$. In the following we will describe in detail how to change these settings. 

\subsection{Observable computation in the SMEFT}
\label{subsec:SMEFT}
In this section we present the main functionalities of the \HighPT program including some code examples, however leaving out some code details necessary, e.g., for creating the plots shown below. The full code examples can be found in the ancillary notebooks available in the GitHub repository~\href{https://github.com/HighPT/HighPT}{\color{\myColor}\faicon{github}}.

The SMEFT run mode is activated by default as discussed before. However, it can also be explicitly specified, e.g.~after reverting back from the mediator mode (see sec.~\ref{subsec:mediator-mode}), by:
\begin{mmaCell}{Input}
  InitializeModel["SMEFT", EFTorder -> 4,
    OperatorDimension -> 6, EFTscale -> 1000]
\end{mmaCell}
\begin{mmaCell}{Print}
  Initialized SMEFT mode:
    Maximum operator dimension: 6
    EFT series truncation at: \mmaSubSup{\(\Lambda\)}{NP}{-4}
    EFT cutoff scale \mmaSub{\(\Lambda\)}{NP}: 1. TeV
\end{mmaCell}
where the option \mmaInlineCell[]{Input}{EFTorder -> 4} determines that cross sections are truncated keeping term up to $\cO(\Lambda^{-4})$ by default.
The option \mmaInlineCell[]{Input}{OperatorDimension -> 6} specifies that by default only operators up to mass-dimension~six should be considered. The default EFT cutoff scale is set to $\Lambda_\mathrm{NP}=1\,\mathrm{TeV}$ by \mmaInlineCell[]{Input}{EFTscale -> 1000}, where one should notice that the unit of energy in \HighPT is~1~GeV. The values listed in this example also correspond to the default choices for \mmaInlineCell[]{Input}{InitializeModel} if any of these options are omitted in the function call.

The \HighPT package contains four main routines that are discussed in more detail below: \mmaInlineCell[]{Input}{DifferentialCrossSection} allows to compute and plot differential hadronic cross sections, \mmaInlineCell[]{Input}{CrossSection} computes the integrated hadronic cross section, \mmaInlineCell[]{Input}{EventYield} derives the predicted number of events in the bins of experimental searches, and \mmaInlineCell[]{Input}{ChiSquareLHC} allows to construct the $\chi^2$~likelihood for the respective searches.
\begin{itemize}
\item \mmaInlineCell[]{Input}{DifferentialCrossSection[\{\mmaUnd{\mmaSub{\(\ell\)}{1}}[\mmaUnd{\(\alpha\)}],\mmaUnd{\mmaSub{\(\ell\)}{2}}[\mmaUnd{\(\beta\)}]\}]}
\\
Computes the differential hadronic cross section $\smash{\frac{\mathrm{d}\sigma}{\mathrm{d}\hat{s}}}$ for the process $pp \to \ell_1^\alpha \bar{\ell}_2^\beta$ in picobarn, which is the default unit for cross sections in \HighPT. The argument \mmaInlineCell[]{Input}{\mmaUnd{\mmaSub{\(\ell\)}{1}}} denotes a lepton, whereas \mmaInlineCell[]{Input}{\mmaUnd{\mmaSub{\(\ell\)}{2}}} denotes an anti-lepton. Thus we require \mmaInlineCell[]{Input}{\mmaUnd{\mmaSub{\(\ell\)}{1}},\mmaUnd{\mmaSub{\(\ell\)}{2}} \(\in\) \{\mmaDef{e},\mmaDef{\(\nu\)}\}} with flavor indices \mmaInlineCell[]{Input}{\mmaUnd{\(\alpha\)},\mmaUnd{\(\beta\)} \(\in\) \{1,2,3\}}. By default the differential cross section is computed by integrating over all values of transverse momentum~$p_T$. The cross section can be restricted to a specific $p_T \in$~\mmaInlineCell[]{Input}{\{\mmaUnd{\mmaSubSup{p}{T}{min}},\mmaUnd{\mmaSubSup{p}{T}{max}}\}} bin using the optional argument \mmaInlineCell[]{Input}{PTcuts -> \{\mmaUnd{\mmaSubSup{p}{T}{min}},\mmaUnd{\mmaSubSup{p}{T}{max}}\}}, where the bin boundaries must be given in units of~GeV. \mmaInlineCell[]{Input}{\mmaDef{DifferentialCrossSection}} returns a function of the partonic center-of-mass energy~$\hat{s}$ that can be evaluated to obtain the differential cross section.
The function \mmaInlineCell[]{Input}{DifferentialCrossSection} takes the following options: \mmaInlineCell[]{Input}{PTcuts}, \mmaInlineCell[]{Input}{FF}, \mmaInlineCell[]{Input}{Coefficients}, \mmaInlineCell[]{Input}{EFTorder}, \mmaInlineCell[]{Input}{OperatorDimension}, and \mmaInlineCell[]{Input}{EFTscale}.
If not explicitly specified, the values given in the \mmaInlineCell[]{Input}{InitializeModel} as default are assumed for the last three options. By default all form factors are substituted by the corresponding Wilson coefficients. This behaviour can be deactivated by specifying the option \mmaInlineCell[]{Input}{FF -> True} which will yield results in terms of the form factors. The option \mmaInlineCell[]{Input}{Coefficients} can be used to specify which Wilson coefficients should be kept in the result, by default all coefficients are included. For further information on these options see the examples below and the additional information provided in tab.~\ref{tab:highpt_options}, as well as the example notebooks.

As an example, we can obtain the differential cross section for $pp \to \mu^- \mu^+$, integrated over the whole $p_T$~range and containing contributions up to $\mathcal{O}(\Lambda^{-4})$, with:
\begin{mmaCell}{Input}
  \mmaDef{\(\sigma\)Dif} = DifferentialCrossSection[\{e[2],e[2]\},
    PTcuts -> \{0,\(\infty\)\}, Coefficients -> \{WC["lq1",\{2,2,1,1\}]\}];
  \mmaDef{\(\sigma\)NP} = \mmaDef{\(\sigma\)Dif} /. WC["lq1",\{2,2,1,1\}] -> 1;
  \mmaDef{\(\sigma\)SM} = \mmaDef{\(\sigma\)Dif} /. WC["lq1",\{2,2,1,1\}] -> 0;
\end{mmaCell}
where we compute the NP contribution~\mmaInlineCell[]{Input}{\mmaDef{\(\sigma\)NP}} for the Wilson coefficient $\big{[}\mathcal{C}_{lq}^{(1)}\big{]}_{2211}\!=1$ by setting all other coefficients to zero. We also compute the SM cross section~\mmaInlineCell[]{Input}{\mmaDef{\(\sigma\)SM}} at tree-level for comparison. The cross sections can then be plotted using:\,\footnote{We omit here and in the following the details about the plotting function for compactness. The full information on how the plots in this section are generated can be found in the example notebooks.} 
\begin{mmaCell}[functionlocal=mll]{Input}
  \mmaDef{LogPlot}[\{\mmaDef{\(\sigma\)NP}[\mmaSup{mll}{2}],\mmaDef{\(\sigma\)SM}[\mmaSup{mll}{2}]\},\{mll,20,600\}]
\end{mmaCell}
\begin{mmaCell}[moregraphics={moreig={scale=1}}]{Output}
  \mmaGraphics{DifferentialCrossSection}
\end{mmaCell}
\item \mmaInlineCell[]{Input}{CrossSection[\{\mmaUnd{\mmaSub{\(\ell\)}{1}}[\mmaUnd{\(\alpha\)}],\mmaUnd{\mmaSub{\(\ell\)}{2}}[\mmaUnd{\(\beta\)}]\}]}
\\
Computes the central value of the integrated hadronic cross section for the process $pp \to \ell_1^\alpha \bar{\ell}_2^\beta$ by using the input from the \mmaInlineCell[]{Input}{\mmaDef{DifferentialCrossSection}} routine. The argument \mmaInlineCell[]{Input}{\mmaUnd{\mmaSub{\(\ell\)}{1}}} denotes a lepton, whereas \mmaInlineCell[]{Input}{\mmaUnd{\mmaSub{\(\ell\)}{2}}} is an anti-lepton. Thus, we require \mmaInlineCell[]{Input}{\mmaUnd{\mmaSub{\(\ell\)}{1}},\mmaUnd{\mmaSub{\(\ell\)}{2}} \(\in\) \{\mmaDef{e},\mmaDef{\(\nu\)}\}} with flavor indices \mmaInlineCell[]{Input}{\mmaUnd{\(\alpha\)},\mmaUnd{\(\beta\)} \(\in\) \{1,2,3\}}. If no flavor index is specified for a neutrino, the summation over all flavors is implicitly understood, i.e.~\mmaInlineCell[]{Input}{CrossSection[\{e[3],\(\nu\)\}} computes the cross section $\sigma(pp\to\tau^-\bar{\nu})=\sum_{i=1}^3\sigma(pp\to\tau^-\bar{\nu}_i)$. By default the cross section is computed by integrating over all values of~$p_T$ and values of~$m_{\ell\ell}$ between $50\,\mathrm{GeV}$ and $13\,\mathrm{TeV}$. The integration boundaries can be modified using the optional arguments \mmaInlineCell[]{Input}{PTcuts -> \{\mmaUnd{\mmaSubSup{p}{T}{min}},\mmaUnd{\mmaSubSup{p}{T}{max}}\}} and \mmaInlineCell[]{Input}{MLLcuts -> \{\mmaUnd{\mmaSubSup{m}{\(\ell\ell\)}{min}},\mmaUnd{\mmaSubSup{m}{\(\ell\ell\)}{max}}\}}, where the bin boundaries must be given in units of~GeV.
The function \mmaInlineCell[]{Input}{CrossSection} takes the following options: \mmaInlineCell[]{Input}{MLLcuts}, \mmaInlineCell[]{Input}{PTcuts}, \mmaInlineCell[]{Input}{FF}, \mmaInlineCell[]{Input}{Coefficients}, \mmaInlineCell[]{Input}{EFTorder}, \mmaInlineCell[]{Input}{OperatorDimension}, and \mmaInlineCell[]{Input}{EFTscale}.

As example, we compute the photon contribution to the SM cross section for $pp \to \mu^- \mu^+$, but this time with $m_{\ell\ell} \in [800\,\mathrm{GeV},2000\,\mathrm{GeV}]$ and without applying any constraint on the transverse momentum. This is done using:
\begin{mmaCell}{Input}
  \mmaDef{\(\sigma\)} = CrossSection[\{e[2],e[2]\},
    MLLcuts -> \{800,2000\}, PTcuts -> \{0,\(\infty\)\},
    FF -> True, Coefficients -> \{FF[Vector,\{"Photon",SM\},_,_]\}];
\end{mmaCell}
Here, we only retain form factors that match the pattern given to \mmaInlineCell[]{Input}{Coefficients}, i.e. \mmaInlineCell[]{Input}{\{FF[Vector,\{"Photon",SM\},_,_]\}}, where the blanks allow for arbitrary values of the chiralities and flavor indices. For more details on the definition of form factors see tab.~\ref{tab:FF_WC}. Thus, in this example we set all form factors to zero except for the SM photon contributions. To find the numerical result, one can use:
\begin{mmaCell}{Input}
  SubstituteFF[\mmaDef{\(\sigma\)}, EFTorder -> 0, OperatorDimension -> 4]
\end{mmaCell}
\begin{mmaCell}{Output}
  0.00947613
\end{mmaCell}
which is given in picobarn, and where the remaining SM form factors have been substituted. The function \mmaInlineCell[]{Input}{SubstituteFF} can also be used to substitute all types of form factors with the corresponding Wilson coefficients or NP coupling constants. For more details see the description in tab.~\ref{tab:highpt}.
\item \mmaInlineCell[]{Input}{EventYield["proc"]}
\\
Computes the expected number of events in each bin of the observable for the experimental search specified by the argument \mmaInlineCell[]{Input}{"proc"}. The available searches are listed in tab.~\ref{tab:searches}, where the second column corresponds to the allowed values of \mmaInlineCell[]{Input}{"proc"}. An overview of these searches can also be obtained using the routine~\mmaInlineCell[]{Input}{LHCSearch[]}.
The event yield is computed by convoluting the corresponding cross sections with the efficiency kernel functions, relating the particle-level observables to the experimental observables (see eq.~\eqref{eq:efficiency_convolution}). These efficiency kernels are obtained by recasting the experimental searches, as described in sec.~\ref{subsec:efficiencies}.
The function prints information about the corresponding experimental analysis and the applied binning in the notebook.\,\footnote{This additionally printed information is omitted in the code examples below for compactness. The printing can be turned off in \cmd{HighPT} by setting \cmd{\$PrintingProcessInfo=False}.} The output is a list where the $n^\text{th}$ entry corresponds to the expected number of events in the $n^\text{th}$ bin of the search.
The function \mmaInlineCell[]{Input}{EventYield} takes the following options: \mmaInlineCell[]{Input}{FF}, \mmaInlineCell[]{Input}{Coefficients}, \mmaInlineCell[]{Input}{EFTorder}, \mmaInlineCell[]{Input}{OperatorDimension}, \mmaInlineCell[]{Input}{EFTscale}, \mmaInlineCell[]{Input}{Luminosity}, \mmaInlineCell[]{Input}{RescaleError}, and \mmaInlineCell[]{Input}{\mmaDef{SM}}. By default (\mmaInlineCell[]{Input}{\mmaDef{SM}->False}) the SM contribution to the event yield is omitted in the output. To include the SM contribution one can specify \mmaInlineCell[]{Input}{\mmaDef{SM}->True}. The options \mmaInlineCell[]{Input}{Luminosity} and \mmaInlineCell[]{Input}{RescaleError} are described below.

For example, we consider the process $pp \to \mu^- \mu^+$ in the respective search by CMS~\cite{CMS:2021ctt}. The relevant observable measured by the experiment is the dilepton invariant mass~$m_\mathrm{\ell\ell}$ of the dimuon system. The expected number of events in each experimental bin for this observable, as a function of only the SMEFT Wilson coefficient~$[\mathcal{C}_{lq}^{(1)}]_{2211}$, including NP squared contributions and assuming $\Lambda=1\,\mathrm{TeV}$, can be obtained by:
\begin{mmaCell}{Input}
  \mmaDef{NEvents} = EventYield["di-muon-CMS",
    Coefficients -> \{WC["lq1",\{2,2,1,1\}]\}];
\end{mmaCell}
For instance, the number of events in the last bin is obtained by:
\begin{mmaCell}{Input}
  \mmaDef{NEvents}[[-1]] // Chop // TraditionalForm
\end{mmaCell}
\begin{mmaCell}{Output}
  - 0.25866 \mmaSub{[\mmaSubSup{C}{lq}{(1)}]}{2211} + 38.9564 \mmaSup{\mmaSub{[\mmaSubSup{C}{lq}{(1)}]}{2211}}{2}
\end{mmaCell}
where the SM contribution is not included.
Notice that one can use \mmaInlineCell[]{Input}{TraditionalForm} to print the output in a more readable format. The event yield can also be exported as a python function using the \cmd{WCxf} format~\cite{Aebischer:2017ugx} by the function call
\begin{mmaCell}{Input}
  PythonExport["di\_muon\_lq1", \mmaDef{NEvents}, "di-muon-CMS"];
\end{mmaCell}
which will create the file \cmd{di\_muon\_lq1.py} in the directory of the notebook. 
This file contains two functions. The \cmd{di\_muon\_CMS\_data()} routine returns a dictionary with keys \cmd{"binning"}, \cmd{"data"}, \cmd{"background"}, and \cmd{"background\_error"}, where each associated value is a list containing the experimental bins, the experimental data per bin, the background per bin as estimated by the experimental collaborations, and the error on the background prediction, respectively.\,\footnote{The background error does not include the statistical uncertainty of the measurement.}
The second function in this file is named \cmd{di\_muon\_lq1(C)}. It takes a dictionary~\cmd{C} of Wilson coefficients in the \cmd{WCxf} format~\cite{Aebischer:2017ugx} as argument and returns a list of the predicted number of events as a functions of the coefficients in each bin.

\item \mmaInlineCell[]{Input}{ChiSquareLHC["proc"]}
\\ 
Constructs the $\chi^2$ likelihood for the observable specified by the string \mmaInlineCell[]{Input}{"proc"}. The supported searches are shown in the tab.~\ref{tab:searches}, where the second column lists all allowed values of the argument~\mmaInlineCell[]{Input}{"proc"}.
An overview of all searches can also be obtained using~\mmaInlineCell[]{Input}{LHCSearch[]}.
\mmaInlineCell[]{Input}{ChiSquareLHC} uses the \mmaInlineCell[]{Input}{EventYield} routine to compute the expectation value for the different bins of the observable in the presence of NP, and the data provided by the experimental collaborations to construct the corresponding $\chi^2$~likelihood for each bin~$k$ defined by
\begin{align}
    \chi^2_k(\theta) &= \frac{(\mathcal{N}_k^\mathrm{NP}(\theta)+\mathcal{N}_k^\mathrm{bkg}-\mathcal{N}_k^\mathrm{obs})^2}{(\Delta\mathcal{N}_k^\mathrm{obs})^2 + (\Delta\mathcal{N}_k^\mathrm{bkg})^2} \,,
    \label{eq:chi2}
\end{align}
where $\theta$ denotes any NP parameter. The signal, background, and observed number of events in the $k^\text{th}$~bin are $\mathcal{N}_k^\mathrm{NP,\,bkg,\,obs}$, respectively. The background uncertainty is $\Delta\mathcal{N}_k^\mathrm{bkg}$, whereas the Poisson uncertainty is $\Delta\mathcal{N}_k^\mathrm{obs}=(\mathcal{N}_k^\mathrm{obs})^{1/2}$.
Information about the experimental search and the binning used for the computation is printed by \mmaInlineCell[]{Input}{ChiSquareLHC} in the notebook when evaluated. \mmaInlineCell[]{Input}{ChiSquareLHC} returns a list where the $k^\text{th}$~element corresponds to the $\chi^2$ for the $k^\text{th}$~bin of the observable.
The function \mmaInlineCell[]{Input}{ChiSquareLHC} takes the following options: \mmaInlineCell[]{Input}{FF}, \mmaInlineCell[]{Input}{Coefficients}, \mmaInlineCell[]{Input}{EFTorder}, \mmaInlineCell[]{Input}{OperatorDimension}, \mmaInlineCell[]{Input}{EFTscale}, \mmaInlineCell[]{Input}{Luminosity}, \mmaInlineCell[]{Input}{RescaleError}, and \mmaInlineCell[]{Input}{CombineBins}. The last option can be used to specify which experimental bins should be combined for the construction of the likelihood. Notice that a combination of bins might be required as a $\chi^2$~likelihood only yields a valid description in the case of sufficiently large number of events. Experimental bins are combined by default such that all resulting bins of the $\chi^2$ contain at least 10~events. Information on the default binning can be obtained by the \mmaInlineCell[]{Input}{LHCSearch["proc"]} routine. In appendix \ref{app:stat} we explicitly checked that with our default binning choice, the exclusion limits extracted with the $\chi^2$ test give comparable results to those extracted using the CL$_s$ method \cite{Read:2000ru}. It is the responsibility of the user to ensure the consistency of the statistical treatment in case they decide to use a finer binning.

As an example, we can compute the $\chi^2$~likelihood for the dimuon search by CMS~\cite{CMS:2021ctt} in the SMEFT keeping the $d=6$ Wilson coefficients $[\mathcal{C}_{lq}^{(1)}]_{2211}$ and $[\mathcal{C}_{lq}^{(1)}]_{2222}$, and terms up to~$\mathcal{O}(\Lambda^{-4})$ in the cross section:
\begin{mmaCell}{Input}
  \mmaDef{L\(\mu\mu\)Binned} = ChiSquareLHC["di-muon-CMS",
    Coefficients -> \{WC["lq1",\{2,2,1,1\}], WC["lq1",\{2,2,2,2\}]\}];
  \mmaDef{L\(\mu\mu\)} = Total[\mmaDef{L\(\mu\mu\)Binned}];
\end{mmaCell}
where \mmaInlineCell[]{Input}{\mmaDef{L\(\mu\mu\)Binned}} is a list with the $k^\mathrm{th}$~element corresponding to the $\chi^2$ of the $k^\mathrm{th}$~bin after merging bins of the experimental search to ensure that each bin has at least 10~events. The combined likelihood \mmaInlineCell[]{Input}{\mmaDef{L\(\mu\mu\)}} can thus be obtained by summing all bins as shown above.
Before deriving the confidence regions for the Wilson coefficients, we also compute the projected likelihood for the high-luminosity run of LHC with an integrated luminosity of $3\,\mathrm{ab}^{-1}$. This is achieved by:
\begin{mmaCell}{Input}
  \mmaDef{L\(\mu\mu\)3000} = Total[ChiSquareLHC["di-muon-CMS",
    Coefficients -> \{WC["lq1",\{2,2,1,1\}], WC["lq1",\{2,2,2,2\}]\},
    Luminosity -> 3000, RescaleError -> True]];
\end{mmaCell}
\begin{mmaCell}{Input}
  \mmaDef{L\(\mu\mu\)3000Const} = Total[ChiSquareLHC["di-muon-CMS",
    Coefficients -> \{WC["lq1",\{2,2,1,1\}], WC["lq1",\{2,2,2,2\}]\},
    Luminosity -> 3000, RescaleError -> False]];
\end{mmaCell}
where \mmaInlineCell[]{Input}{\mmaDef{L\(\mu\mu\)3000}} is the combined likelihood projected to $3\,\mathrm{ab}^{-1}$ assuming that the background uncertainty is rescaled by $\Delta\mathcal{N}_k^\mathrm{bkg} \rightarrow (L_\mathrm{projected}/L_\mathrm{current})^{1/2} \,\Delta\mathcal{N}_k^\mathrm{bkg}=\sqrt{3000/140}\,\Delta\mathcal{N}_k^\mathrm{bkg}$. On the contrary, \mmaInlineCell[]{Input}{\mmaDef{L\(\mu\mu\)3000Const}} is the likelihood computed assuming that the ratio of background error over background is constant, i.e. $\Delta\mathcal{N}_k^\mathrm{bkg}/\mathcal{N}_k^\mathrm{bkg}=\text{const}$.
We can then minimize the resulting likelihoods to plot the 95\% CL contours for the two specified Wilson coefficients:
\begin{mmaCell}[moregraphics={moreig={scale=0.5}}]{Output}
  \mmaGraphics{dimuon_projection}
\end{mmaCell}
The blue region represent the current $95\%\,\mathrm{CL}$ region for the coefficients, whereas the orange and green regions show the projection for a luminosity of $3\,\mathrm{ab}^{-1}$ assuming the background uncertainty is rescaled by a factor $\smash{(L_\mathrm{projected}/L_\mathrm{current})^{1/2}}$ and 
$L_\mathrm{projected}/L_\mathrm{current}$, respectively, where $L_\mathrm{current}=140\,\mathrm{fb}^{-1}$ and $L_\mathrm{projected}=3000\,\mathrm{fb}^{-1}$.
\end{itemize}

\paragraph{CKM matrix and mass-basis alignment}
By default \HighPT uses a non-diagonal CKM matrix and assumes down alignment, i.e. $V_d=\mathbb{1}$ and $V_u=V_\mathrm{CKM}^\dagger$ as defined in eq.~\eqref{eq:qL_rotation_matrices}. However, any alignment can be specified with the function \mmaInlineCell[]{Input}{DefineBasisAlignment[arg]}, where \mmaInlineCell[]{Input}{arg} can be \mmaInlineCell[]{Input}{"up"} or \mmaInlineCell[]{Input}{"down"} to specify up alignment or down alignment, respectively. Alternatively, any $3 \times 3$~matrix can be given as \mmaInlineCell[]{Input}{arg}, in which case \HighPT defines $V_d=$\,\mmaInlineCell[]{Input}{arg} and $V_u=V_d \cdot V_\mathrm{CKM}^\dagger$. The current value of the CKM matrix elements can be obtained using the \mmaInlineCell[]{Input}{GetParameters[]} function and their values can be modified using the routine \mmaInlineCell[]{Input}{DefineParameters["Wolfenstein"->\{\mmaUnd{\(\lambda\)},A,\mmaUnd{\mmaOver{\(\rho\)}{_}},\mmaUnd{\mmaOver{\(\eta\)}{_}}\}]}, where \mmaInlineCell[]{Input}{\mmaUnd{\(\lambda\)},A,\mmaUnd{\mmaOver{\(\rho\)}{_}},\mmaUnd{\mmaOver{\(\eta\)}{_}}} specify the Wolfenstein parameters.
As an example we derive bounds on the Wilson coefficients of scalar operators for second-generation fermions assuming up alignment, down alignment, and a diagonal CKM matrix, respectively. The corresponding likelihoods can be obtained by:
\begin{mmaCell}{Input}
  \mmaDef{wc} = \{WC["ledq",\{2,2,2,2\}], WC["lequ1",\{2,2,2,2\}]\};
  DefineBasisAlignment["up"];
  \mmaDef{Lup} = Total[ChiSquareLHC["di-muon-CMS", Coefficients->\mmaDef{wc}]];
  DefineBasisAlignment["down"];
  \mmaDef{Ldown} = Total[ChiSquareLHC["di-muon-CMS", Coefficients->\mmaDef{wc}]];
  DefineParameters["Wolfenstein"->\{0,0,0,0\}];
  \mmaDef{Lno} = Total[ChiSquareLHC["di-muon-CMS", Coefficients->\mmaDef{wc}]];
  DefineParameters[Default];
\end{mmaCell}
where in the last line we set the CKM back to its default value. Minimizing and plotting the likelihoods thus obtained we find:
\begin{mmaCell}[moregraphics={moreig={scale=0.5}}]{Output}
  \mmaGraphics{basis_alignment.png}
\end{mmaCell}
In the case of a diagonal $V_\mathrm{CKM}$ (black dashed line) we find that $\smash{[C_{ledq}]_{2222}}$ is more strongly constrained than $\smash{[C_{lequ}^{(1)}]_{2222}}$ due to the strange PDF dominating over the charm PDF. Using a non-diagonal CKM and assuming down alignment (blue region), we see that the constraint on $\smash{[C_{lequ}^{(1)}]_{2222}}$ becomes stronger due to the additional contribution from the up-quark PDF that is only mildly Cabibbo suppressed. For the up-aligned case (yellow region), we find that the constraint on $\smash{[C_{ledq}]_{2222}}$ improves due to the down-PDF contribution with similar Cabibbo suppression. However, due to the suppression of the down PDF with respect to the up PDF, this effect is less pronounced than in the down-aligned scenario. For further discussion of the basis alignment see~ref.~\cite{Allwicher:2022}.

\subsection{Specific NP models with BSM mediators}
\label{subsec:mediator-mode}
In addition to computing cross sections and likelihoods within the SMEFT, \HighPT also allows to perform such computations in concrete NP models with explicit heavy bosonic tree-level mediators. A~list of all currently supported mediators and their interaction Lagrangians is given in tab.~\ref{tab:mediators}. 
As of version~\cmd{1.0.0}, \HighPT will support only a few selected masses for the BSM mediators.\,\footnote{An updated list of currently available mediator masses can be found at~\href{https://github.com/HighPT/HighPT}{\color{\myColor}\faicon{github}}.}${}^{,}$\footnote{Notice moreover that the Monte Carlo simulations have been generated assuming zero width for the leptoquarks. A non-zero width can however be specified in the code (see below), the zero-width efficiencies will then be used for the computation. The error committed by doing so is negligible for $t$- and $u$-channel exchanges.}
Multiple mediators can be considered at the same time, however the interference terms of two BSM mediators are not computed for event yields and likelihoods due to missing Monte Carlo simulations. Further masses and the interference terms will be included in the future.

\begin{table}[h!]

  \begin{center}
  {\renewcommand{\arraystretch}{1.4}
    \begin{tabular}{@{\hspace{1em}}c@{\hspace{1em}}c@{\hspace{1em}}c@{\hspace{1em}}c@{\hspace{1em}}l@{\hspace{1em}}}
    \ & \HighPT label &   SM rep.  & Spin  & \hspace{3cm} $\cL_{\rm int}$\\
\hline\hline
$Z^\prime$ & \mmaInlineCell[]{Input}{"Zp"} & \SMrep{1}{1}{0} & 1 & $\cL_{Z^\prime}=\sum_\psi\, [g^{\psi}_1]_{ab} \, \bar \psi_a \slashed Z^\prime \psi_b$\, , \  $\psi\in\{u,d,e,q,l\}$\\
$W^\prime$ & \mmaInlineCell[]{Input}{"Wp"} & \SMrep{1}{3}{0} & 1 & $\cL_{W^\prime}=[g^{q}_3]_{ij} \, \bar q_i (\tau^I \,\slashed W^{\prime\, I}) q_j+[g^{l}_3]_{\alpha\beta} \, \bar l_\alpha (\tau^I \,\slashed W^{\prime\, I}) l_\beta$ \\
\hline
$S_1$ & \mmaInlineCell[]{Input}{"S1"} & $\SMrepbar{3}{1}{1/3}$  & 0 & $\cL_{S_1}= [y_1^L]_{i\alpha}\, S_1 \bar  q^c_i\epsilon l_\alpha+[y_1^R]_{i\alpha}\,S_1\bar u^c_i e_\alpha +\mathrm{h.c.}$\\ 
$\widetilde S_1$ & \mmaInlineCell[]{Input}{"S1t"} & $\SMrepbar{3}{1}{4/3}$ & 0 & $\cL_{\widetilde S_1}=[\widetilde y_1^R]_{i\alpha}\,  \widetilde S_1\bar d^c_i e_\alpha+\mathrm{h.c.}$\\ 
$U_1$ & \mmaInlineCell[]{Input}{"U1"} & $\SMrep{3}{1}{2/3}$ & 1 & $\cL_{U_1}=[x_1^L]_{i\alpha} \, \bar q_i \slashed  U_{\!1} l_\alpha + [x_1^R]_{i\alpha} \,\bar d_i \slashed U_{\!1} e_\alpha + \mathrm{h.c.}$\\ 
$\widetilde U_1$ & \mmaInlineCell[]{Input}{"U1t"} & $$\SMrep{3}{1}{5/3}$$ & 1 & $\cL_{\widetilde U_1}= [\widetilde x_1^R]_{i\alpha} \,\bar u_i \slashed {\widetilde U}_{\!1} e_\alpha+\mathrm{h.c.}$\\ 
$R_2$ & \mmaInlineCell[]{Input}{"R2"} & \SMrep{3}{2}{7/6} & 0  & $\cL_{R_2}= -[y_2^L]_{i\alpha} \, \bar u_i R_2 \epsilon l_\alpha+[y_2^R]_{i\alpha} \, \bar q_i e_\alpha R_2+\mathrm{h.c.}$ \\ 
$\widetilde R_2$ & \mmaInlineCell[]{Input}{"R2t"} & \SMrep{3}{2}{1/6} & 0 & $\cL_{\widetilde R_2}=- [\widetilde y_2^L]_{i\alpha} \, \bar d_i \widetilde R_2 \epsilon l_\alpha+ \mathrm{h.c.}$ \\ 
$V_2$ & \mmaInlineCell[]{Input}{"V2"} & $\SMrepbar{3}{2}{5/6}$ & 1 & $\cL_{V_{\!2}}=[x_2^L]_{i\alpha} \,\bar d_i^c \slashed V_{\!\!2}\epsilon l_\alpha+[x_2^R]_{i\alpha}\, \bar q_i^c\epsilon \slashed V_{\!\!2} e_\alpha +\mathrm{h.c.}$\\ 
$\widetilde V_2$ & \mmaInlineCell[]{Input}{"V2t"} & $\SMrepbar{3}{2}{-1/6}$  & 1 & $\cL_{\widetilde V_{\!2}}=[\widetilde x_2^L]_{i\alpha} \,\bar u_i^c \slashed {\widetilde V}_{\!\!2}\epsilon l_\alpha+ \mathrm{h.c.}$\\ 
$S_3$ & \mmaInlineCell[]{Input}{"S3"} & $\SMrepbar{3}{3}{1/3}$ & 0 & $\cL_{S_3}=[y_3^L]_{i\alpha} \, \bar q^c_i \epsilon (\tau^I \, S_3^I) l_{\alpha}+\mathrm{h.c.}$\\ 
$ U_3$ & \mmaInlineCell[]{Input}{"U3"} & $\SMrep{3}{3}{2/3}$ & 1 & $\cL_{U_3}=[x_3^L]_{i\alpha} \, \bar q_i (\tau^I \, \slashed {U}_{\!3}^I ) l_\alpha+\mathrm{h.c.}$ \\
\hline\hline
\end{tabular}
}
\caption{Bosonic mediators contributing at tree level to Drell-Yan production which are implemented in {\HighPT} are classified by their SM quantum numbers and spin. In the last column, we provide the interaction Lagrangian where $\epsilon\equiv i\tau_2$, $\psi^c\equiv i\gamma_2\gamma_0\bar\psi^T$, where $\tau_i$ ($i=1,2,3$) denote the Pauli matrices. Right-handed fermion fields are denoted by $u\equiv u_R$, $d\equiv d_R$, $e\equiv \ell_R$ and left-handed fermion fields by $q\equiv (u_L,d_L)^T$, $l\equiv (\nu_L,\ell_L)^T$. For the leptoquark states we adopt the symbols from ref.~\cite{Dorsner:2016wpm}.}
\label{tab:mediators}
\end{center}
\end{table}
To define a model with specific mediators, the \mmaInlineCell[]{Input}{InitializeModel} routine can be used as:
\begin{mmaCell}{Input}
  InitializeModel["Mediators", 
    Mediators -> \{"label1" -> \{mass1,width1\},...\}]
\end{mmaCell}
where \mmaInlineCell[]{Input}{"label1"} is the label of the first mediator as listed in the second column of tab.~\ref{tab:mediators} and \mmaInlineCell[]{Input}{mass1} (\mmaInlineCell[]{Input}{width1}) is its mass (width) given in units of~GeV. The ellipsis denote possible further mediators defined in the same way. In this case all EFT interactions are turned off. All functions for computing cross sections or likelihoods can be used in the same way as in the EFT scenario. To revert back to the SMEFT run mode one can use \mmaInlineCell[]{Input}{InitializeModel["SMEFT"]} as discussed in sec.~\ref{subsec:SMEFT}.

As an example, we can derive the likelihood for a model with a $2\,\mathrm{TeV}$ $U_1$~vector leptoquark and a $3\,\mathrm{TeV}$ $R_2$~scalar leptoquark which both have zero width in approximation. We analyze the contribution in this model to $pp \to \tau\tau$ and $pp \to \tau\nu$ by:
\begin{mmaCell}{Input}
  InitializeModel["Mediators", 
    Mediators -> \{"U1" -> \{2000, 0\}, "R2" -> \{3000, 0\}\}];
\end{mmaCell}
\begin{mmaCell}{Input}
  \mmaDef{\(\chi\)SqLQ\(\tau\tau\)} = Total[ChiSquareLHC["di-tau-ATLAS"]];
  \mmaDef{\(\chi\)SqLQ\(\tau\nu\)} = Total[ChiSquareLHC["mono-tau-ATLAS"]];
\end{mmaCell}
Here we directly sum the $\chi^2$~likelihoods for each individual bin of the experimental search by applying \mmaInlineCell[]{Input}{Total}. By default the form factors are substituted by the NP couplings to fermions.
We can now perform a fit to the couplings constants~$[x_1^L]_{13}$ and~$[y_2^L]_{13}$ while neglecting all other couplings, obtaining:
\begin{mmaCell}[moregraphics={moreig={scale=0.5}}]{Output}
  \mmaGraphics{LQ}
\end{mmaCell}
Notice that the $U_1$ and $R_2$ leptoquarks do not interfere. Therefore, the full likelihood is computed above. For interfering mediators, the interference of the BSM mediators cannot be computed due to missing Monte Carlo simulations and they are thus set to zero.

The other routines such as \mmaInlineCell[]{Input}{DifferentialCrossSection}, \mmaInlineCell[]{Input}{CrossSection}, and \mmaInlineCell[]{Input}{EventYield} can be used in the mediator mode of \HighPT in the same way as in the SMEFT mode.
The optional arguments \mmaInlineCell[]{Input}{EFTorder}, \mmaInlineCell[]{Input}{OperatorDimension}, and \mmaInlineCell[]{Input}{EFTscale} are ignored in this case.

\section{NP constraints from high-$p_T$ tails}
\label{sec:examples}
After having introduced the main functionalities of \HighPT, we will now consider more detailed examples. We start in sec.~\ref{subsec:EFT-example} with an illustration of how to constrain SMEFT Wilson coefficients at different orders in the EFT power counting and with a comparison of these results. In sec.~\ref{subsec:LFV-example}, we consider a model with a $S_3$~scalar leptoquark that leads to Lepton Flavor (Universality) Violating (LF(U)V) effects and investigate the constraints that can be derived on the respective couplings. The full code for the examples of this section can be found in the ancillary \Mathematica notebooks.

\subsection{Comparing limits at different EFT orders}
\label{subsec:EFT-example}
\begin{figure}
    \centering
    \includegraphics[width=0.49\textwidth]{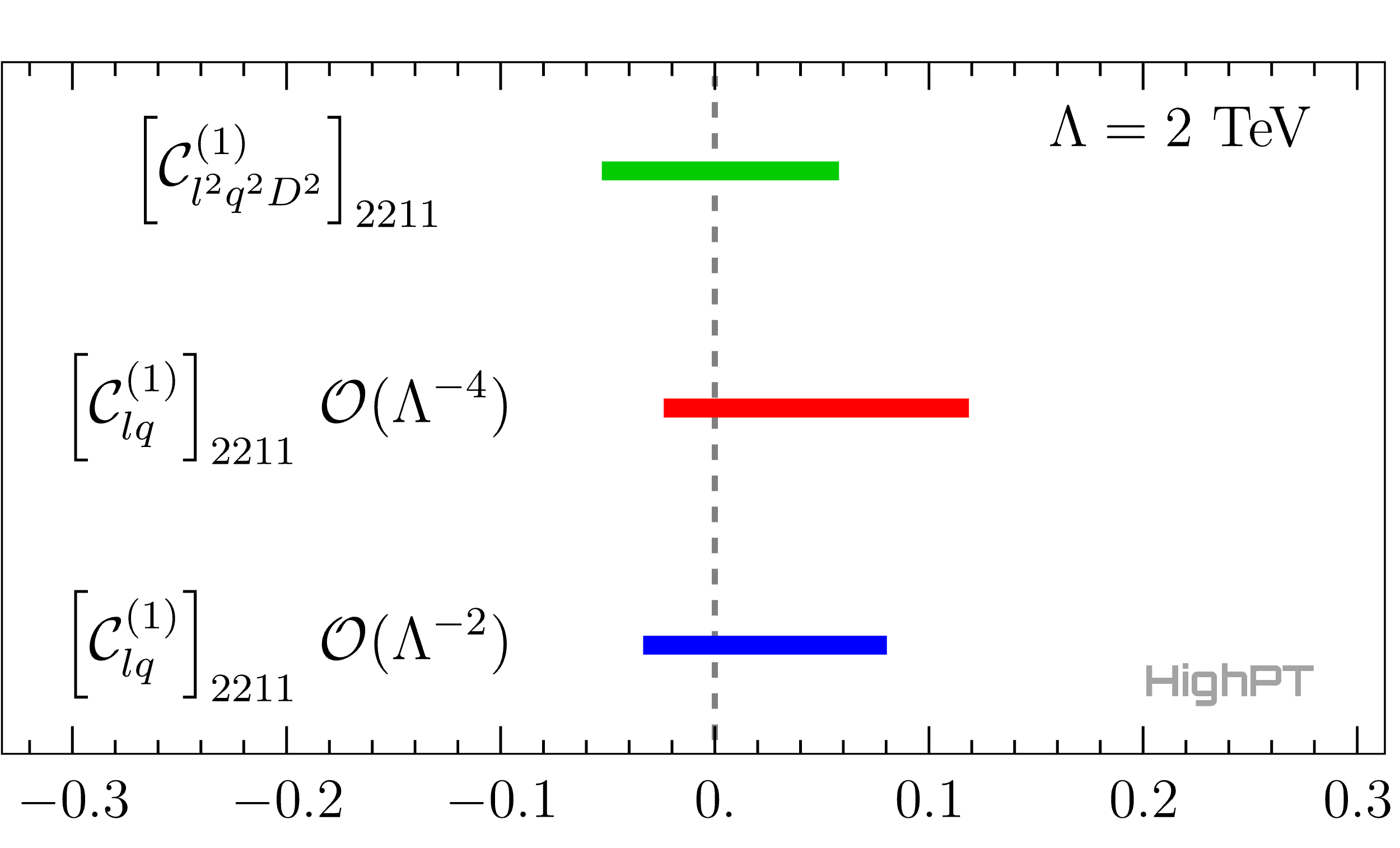}
    \includegraphics[width=0.49\textwidth]{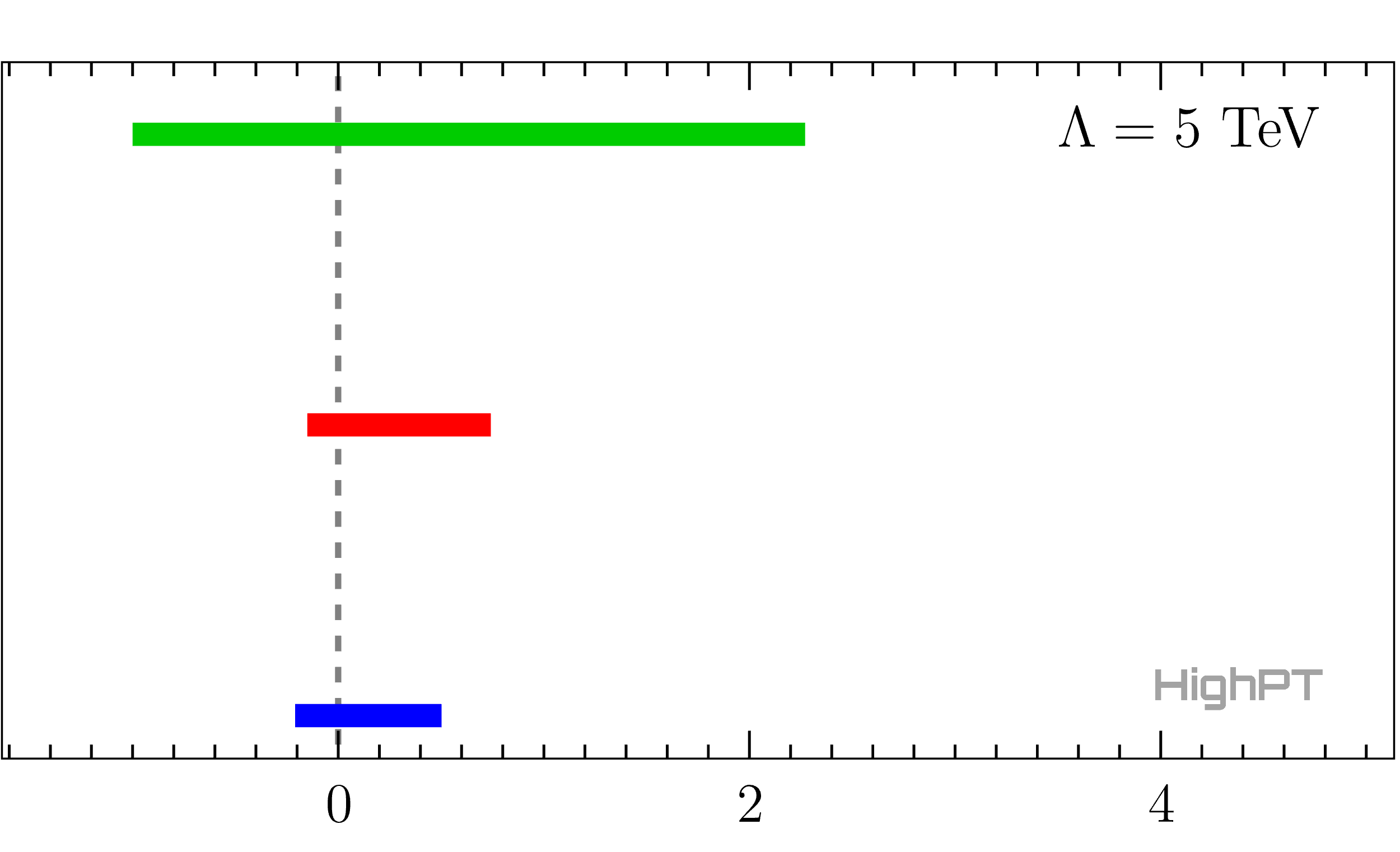}
    \caption{$2\sigma$ confidence intervals for different single Wilson coefficient scenarios at different orders in the EFT expansion.}
    \label{fig:SMEFTexample}
\end{figure}

In this section we show how to truncate the EFT series on different powers of $\Lambda$ and how to treat $d=8$ operators in \HighPT.
As a first step, we construct the $\chi^2$ truncating the cross section at~$\mathcal{O}(\Lambda^{-2})$. We use the CMS dimuon search, and choose to switch on only the operator $\smash{[\cO_{lq}^{(1)}]_{2211}}$. Remember that the routine \mmaInlineCell[]{Input}{InitializeModel} sets the default values for the options \mmaInlineCell[]{Input}{EFTorder} and \mmaInlineCell[]{Input}{OperatorDimension}. These values will have to be overwritten in order to obtain the remaining $\chi^2$ functions in this section:
\begin{mmaCell}{Input}
  InitializeModel["SMEFT", EFTorder->2, 
    OperatorDimension->6, EFTscale->2000];
\end{mmaCell}
\begin{mmaCell}{Input}
  \mmaDef{L\(\mu\mu\)6int} = ChiSquareLHC["di-muon-CMS", 
    Coefficients->\{WC["lq1",\{2,2,1,1\}]\}] // Total;
\end{mmaCell}
Constructing the $\chi^2$ by truncating the cross section at~$\mathcal{O}(\Lambda^{-4})$ with only $d=6$ operators is possible by using:
\begin{mmaCell}{Input}
  \mmaDef{L\(\mu\mu\)6sq} = ChiSquareLHC["di-muon-CMS", 
    EFTorder->4, OperatorDimension->6, 
    Coefficients->\{WC["lq1",\{2,2,1,1\}]\}] // Total;
\end{mmaCell}
The $\chi^2$ distribution with a truncation on the cross section at~$\mathcal{O}(\Lambda^{-4})$, keeping only the interference of the $d=8$ terms with the SM,\,\footnote{In general these settings would give results containing both $d=6$ and $d=8$ operators. However, here we explicitly specify that only one $d=8$ coefficient should be kept.} can be obtained via:
\begin{mmaCell}{Input}
  \mmaDef{L\(\mu\mu\)8int} = ChiSquareLHC["di-muon-CMS", 
    EFTorder->4, OperatorDimension->8, 
    Coefficients->\{WC["l2q2D21",\{2,2,1,1\}]\}] // Total;
\end{mmaCell}
Following a similar procedure to the examples above, we can now compare the fits in the three cases in fig.~\ref{fig:SMEFTexample}.
We see that, for $\Lambda = 2$ TeV, the sensitivity to the $d=8$ coefficient is quite good, having a similar bound coming from the $d=6$ interference with the SM. This is expected since the two contributions differ by a factor $\hat s/\Lambda^2$, which is $\mathcal{O}(1)$ in this case.
On the other hand, for $\Lambda = 5\,\mathrm{TeV}$, the effect from dimension eight becomes less relevant.

\subsection{Constraining LF(U)V in leptoquark models}
\label{subsec:LFV-example}
As a last example, we consider the $S_3\sim ({\bf \bar{3}}, {\bf 3}, {1/3})$ scalar leptoquark,
\begin{equation}
\cL_{S_3}=[y_3^L]_{i\alpha} \, \bar q^c_i \epsilon (\tau^I \, S_3^I) l_{\alpha}+\mathrm{h.c.}\,,
\end{equation}
where we assume that only the couplings to third generation quarks $[y_3^L]_{3\alpha}$ and $[y_3^L]_{3\beta}$ are non-vanishing. This is an illustrative example as the leptoquark will simultaneously contribute to the LFV process $pp\to\ell_\alpha\ell_\beta$ (with $\alpha\neq\beta$), in addition to the ones that conserve lepton flavor, namely $pp\to\ell_\alpha\ell_\alpha$ and $pp\to\ell_\beta\ell_\beta$.\,\footnote{The contribution to monolepton final states is CKM suppressed in this example and is therefore omitted.} Firstly, we initialize the $S_3$~model in \HighPT and load the relevant searches:

\begin{figure}[!t]
    \centering
    \includegraphics[width=0.32\textwidth]{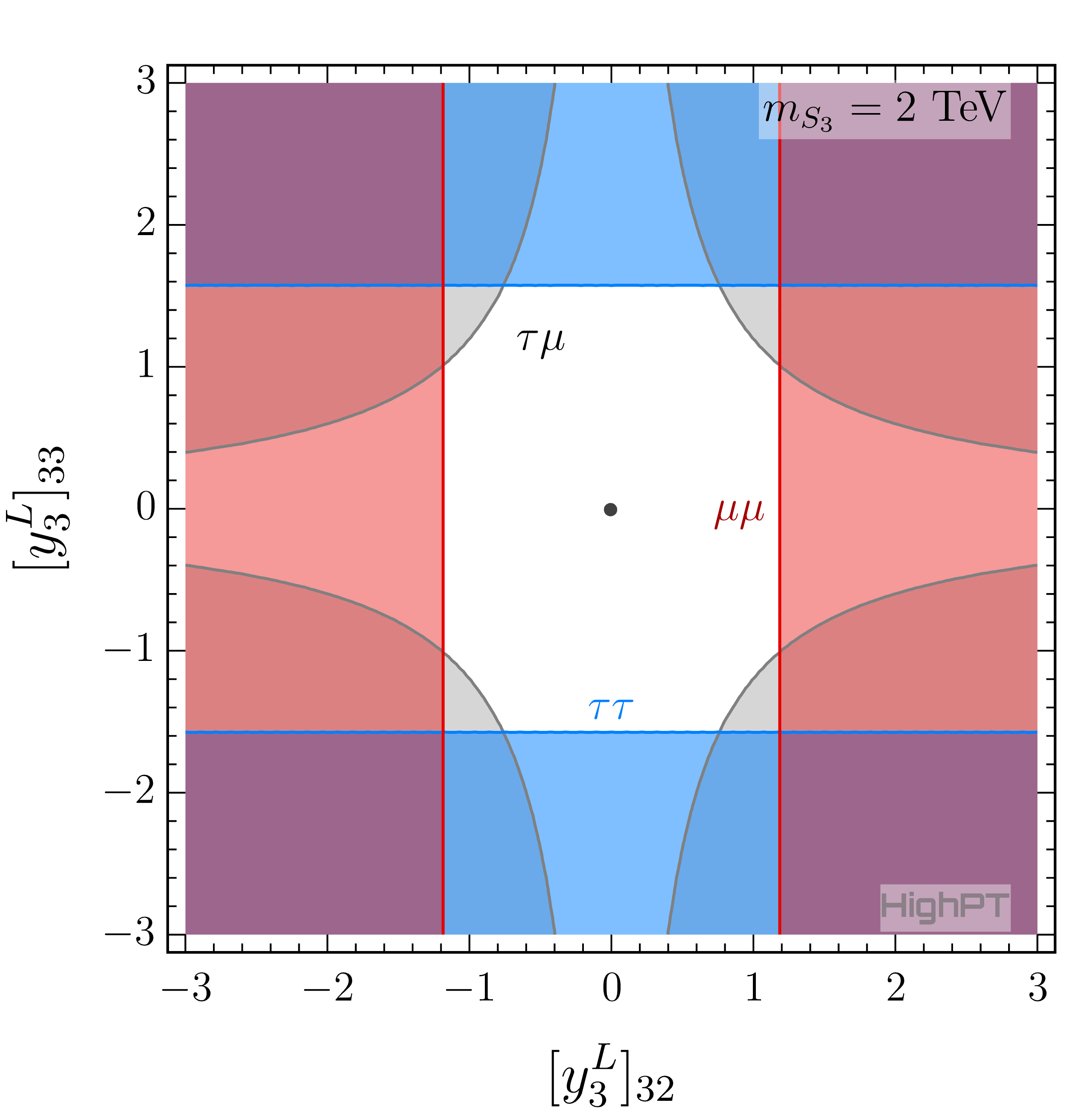}
    \includegraphics[width=0.32\textwidth]{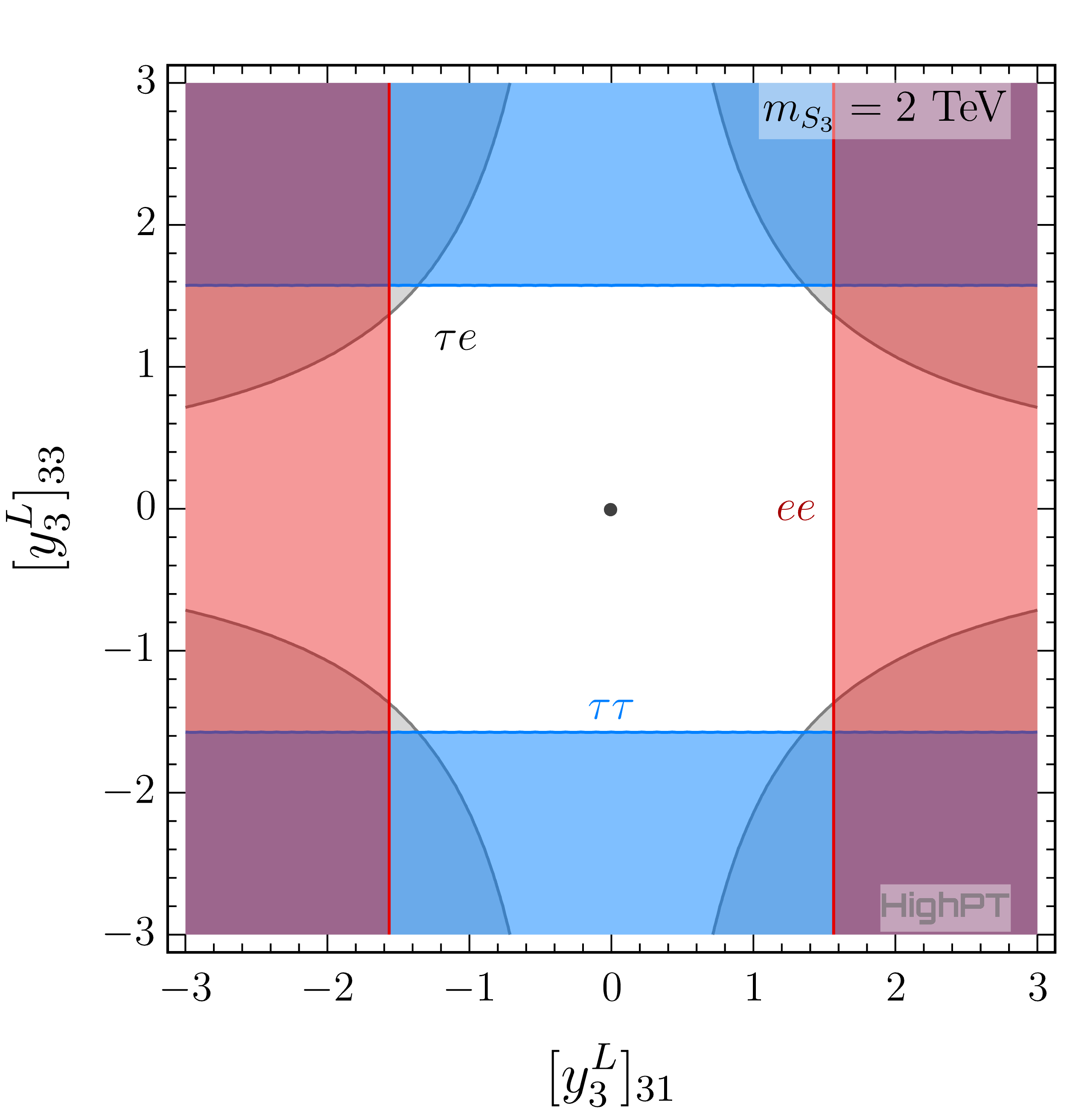}
    \includegraphics[width=0.32\textwidth]{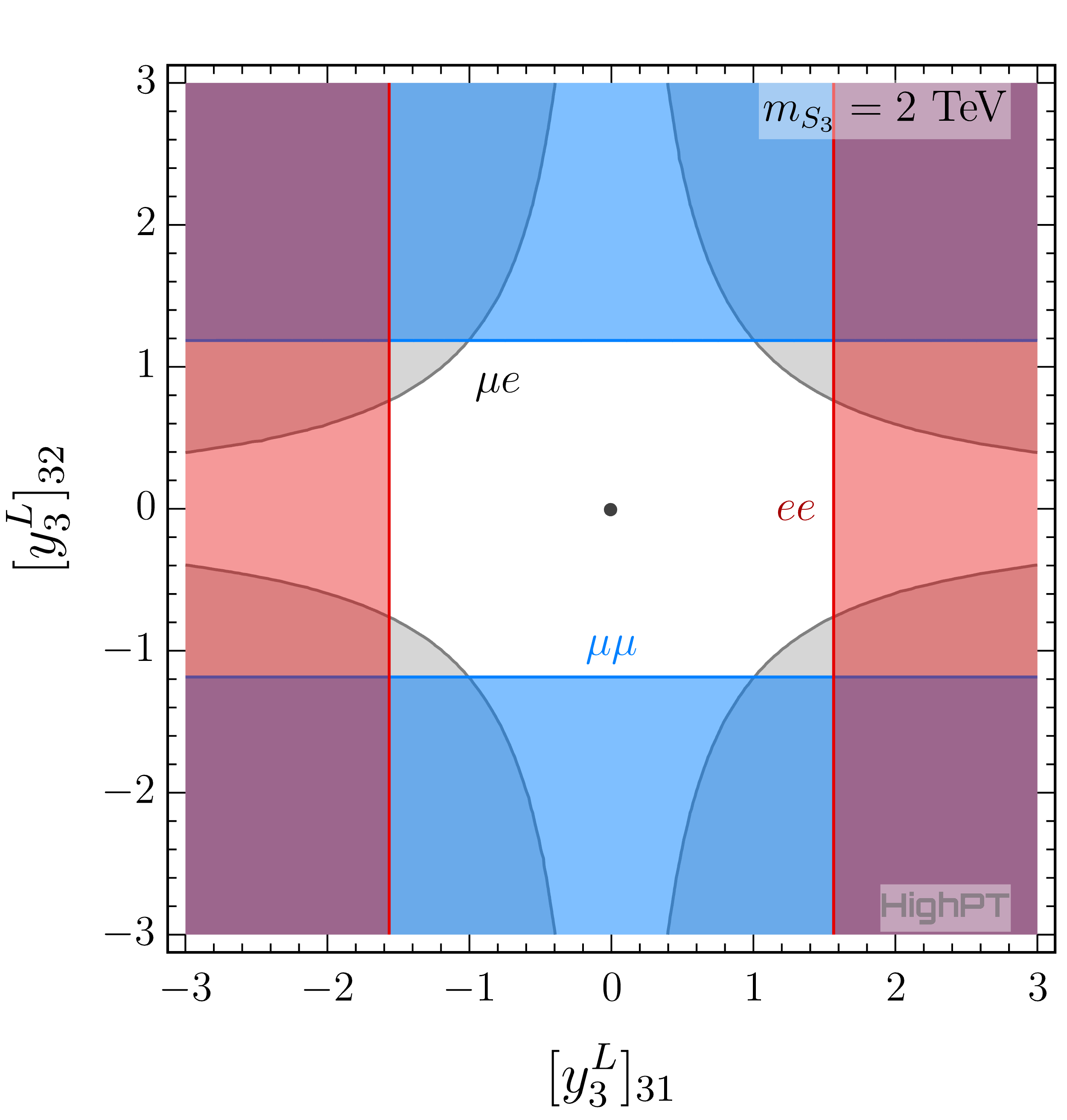}
    \caption{Two-dimensional exclusion regions for the left-handed $S_3$ couplings. All contours indicate $2\sigma$ exclusion.}
    \label{fig:LFV}
\end{figure}

\begin{mmaCell}{Input}
  InitializeModel["Mediators", Mediators -> \{"S3" -> \{2000, 0\}\}];
  \mmaDef{L\(\mu\tau\)} = ChiSquareLHC["muon-tau-CMS"] // Total;
  \mmaDef{Le\(\tau\)} = ChiSquareLHC["electron-tau-CMS"] // Total;
  \mmaDef{Le\(\mu\)} = ChiSquareLHC["electron-muon-CMS"] // Total;
  \mmaDef{L\(\tau\tau\)} = ChiSquareLHC["di-tau-ATLAS"] // Total;
  \mmaDef{L\(\mu\mu\)} = ChiSquareLHC["di-muon-CMS"] // Total;
  \mmaDef{Lee} = ChiSquareLHC["di-electron-CMS"] // Total;
\end{mmaCell}
where we immediately combined the likelihoods of all bins for each search. These likelihoods can then be used to plot the allowed couplings on different planes, as depicted in fig.~\ref{fig:LFV} for $m_{S_3}=2\,\mathrm{TeV}$. Lepton flavor conserving searches can separately constrain the couplings $[y_3^L]_{3\alpha}$ and $[y_3^L]_{3\beta}$ (with $\alpha\neq \beta$), implying an indirect limit on $[y_3^L]_{3\alpha}\,[y_3^L]_{3\beta}$. The latter combination of couplings enters directly the cross section of the LFV process $pp\to \ell_\alpha^\pm\ell_\beta^\mp$. From fig.~\ref{fig:LFV}, we see that there is an important complementarity between the lepton flavor conserving and violating searches.

\section{Conclusion and Outlook}\label{sec:conclusion}

We have introduced the \Mathematica package \HighPT, a self-contained tool for performing high-$p_T$ tail analyses of various New Physics scenarios at the LHC. The user can use
HighPT to easily perform analyses of various New Physics scenarios at the LHC. With \HighPT, the user can easily compute the hadronic (differential) cross sections for monolepton and dilepton production at hadron colliders, and construct a $\chi^2$~likelihood through a recast of the most recent experimental searches using the LHC run-II data-sets. Currently, the complete likelihoods can be obtained both for the SMEFT up to $d\leq 8$ and for selected simplified scenarios, defined by one or more bosonic mediators in addition to the SM Lagrangian. See~\cite{Allwicher:2022} for a detailed phenomenological analysis using \HighPT.  

This first release of \HighPT represents just the initial step towards a more complete and flexible environment. In particular, the following features will be added in future releases: 

\begin{itemize}

    \item Implementation of low-energy flavor observables, as well as  electroweak and Higgs data, including the relevant running effects and the matching between the low-energy effective theory and the SMEFT, in order to be able to extract the full combined likelihood in one framework.  
    
    \item So far, the mediator mode is only available for leptoquarks and a few selected masses. This will be improved in the future with a finer scan. Furthermore, we also plan to include possible interference effects when analyzing several leptoquark mediators at a time.
    
    \item Currently, the package only supports single momentum-dependent observables such as the dilepton invariant mass for $pp\to\ell\ell$ or the lepton transverse mass for $pp\to\ell\nu$. In the future we plan to include LHC searches measuring double-differential distributions and angular-based observables, such as the forward-backward asymmetry.   
    
    \item Going beyond the SMEFT and leptoquarks, we plan to include the full likelihoods for any $s$-channel mediator, as well as effective operators with light right-handed neutrinos.  
    
    \item Alternative statistical treatments like the $\mathrm{CL}_s$ method~\cite{Read:2002hq}, possibly interfacing with the \cmd{pyhf} package~\cite{Heinrich:2021gyp}, see appendix~\ref{app:stat}.
    
    \item To allow the user to include various sources of theoretical uncertainties, e.g.~higher order QCD/EW corrections~\cite{Degrande:2020evl} and PDF uncertainties.
    
    \item Any sufficiently heavy New Physics can be matched onto the SMEFT. This matching can be performed with existing automatized tools like \cmd{MatchMakerEFT}~\cite{Carmona:2021xtq} and \cmd{Matchete}~\cite{Matchete}, allowing in principle for a fully automated pipeline to get constraints on the ultraviolet models.
\end{itemize}

\section*{Acknowledgements}
We sincerely thank Gino Isidori for his encouragement to pursue this project. We thank Jason Aebischer for discussions regarding the \cmd{WCxf} format.
This project has received funding from the European Research Council (ERC) under the European Union's Horizon 2020 research and innovation programme under the grant agreement 833280 (FLAY) and the Marie Skłodowska-Curie grant agreement No~860881-HIDDeN, and by the Swiss National Science Foundation (SNF) under contract 200020-204428.

\appendix
\section{Overview of all HighPT variables}
\label{app:functions}
In this appendix we list global variables used in \HighPT in tab.~\ref{tab:highpt}~--~\ref{tab:highpt_options}. Together with the main routines, which where already discussed before in sec.~\ref{sec:HighPT}, the tables given below provide a complete list of all global variables available as of version~\cmd{1.0.0}.
\begin{center}
\begin{longtable}{| p{.35\textwidth} | p{.65\textwidth} |} 
\hline
\textbf{Routines} & \textbf{Description}
\\\hline
\mmaInlineCell[]{Input}{LHCSearch[]}
&
Lists all the experimental searches available in \HighPT. Detailed information on one particular search, e.g., the ditau search by ATLAS can be obtained using \mmaInlineCell[]{Input}{LHCSearch["di-tau-ATLAS"]}.
\\\hline
\mmaInlineCell[]{Input}{\mmaDef{SubstituteFF}[\mmaUnd{expression}]}
&
If running in the SMEFT mode \mmaInlineCell[]{Input}{SubstituteFF} maps the \mmaInlineCell[]{Input}{\mmaUnd{expression}} given in terms of form factors~(\mmaInlineCell[]{Input}{\mmaDef{FF}}) to the SMEFT Wilson coefficients~(\mmaInlineCell[]{Input}{\mmaDef{WC}}) in the $d=6$ Warsaw basis~\cite{Grzadkowski:2010es} and the $d=8$~basis given in ref.~\cite{Murphy:2020rsh}. The maximum mass dimension~\mmaInlineCell[]{Input}{\mmaUnd{d}} of EFT operators to consider can be modified by specifying the option \mmaInlineCell[]{Input}{\mmaDef{OperatorDimension}->\mmaUnd{d}}, where \mmaInlineCell[]{Input}{\mmaUnd{d}}$\in\{4,6,8\}$. The truncation of the EFT series up to and including terms of order~$\cO(\Lambda$\mmaInlineCell[]{Input}{\mmaSup{}{-n}}$)$ can be specified using \mmaInlineCell[]{Input}{\mmaDef{EFTorder}->n}, where \mmaInlineCell[]{Input}{\mmaUnd{n}}$\in\{0,2,4\}$. The EFT cutoff scale~\mmaInlineCell[]{Input}{\mmaUnd{\(\Lambda\)}} can be specified (in units of GeV)  by \mmaInlineCell[]{Input}{EFTscale->\mmaUnd{\(\Lambda\)}}. If the above options are not given, the default values, as defined in the last call of the \mmaInlineCell[]{Input}{InitializeModel} routine, are assumed. Notice that applying \mmaInlineCell[]{Input}{SubstituteFF} to a likelihood expressed in terms of form factors will yield inconsistent results, as the EFT power series must be truncated on observable level, i.e.~for the cross section and not the likelihood. Thus, \mmaInlineCell[]{Input}{ChiSquareLHC} automatically truncates on cross section level when outputting results in terms of Wilson coefficients.

If running in the mediator mode, all form factors in \mmaInlineCell[]{Input}{\mmaUnd{expression}} are replaced by the appropriate NP couplings to fermions (\mmaInlineCell[]{Input}{Coupling}) of the model that was specified in the last call of \mmaInlineCell[]{Input}{InitializeModel}. 
All options mentioned above are ignored in this case.
\\\hline
\mmaInlineCell[]{Input}{SelectTerms[expr,list]}
&
Takes a \mmaInlineCell[]{Input}{list} of Wilson coefficients~(\mmaInlineCell[]{Input}{WC}), couplings~(\mmaInlineCell[]{Input}{Coupling}) and/or form factors~(\mmaInlineCell[]{Input}{FF}), and sets all appearances of \mmaInlineCell[]{Input}{WC}, \mmaInlineCell[]{Input}{Coupling} and \mmaInlineCell[]{Input}{FF} in~\mmaInlineCell[]{Input}{expr} that do not match the patterns specified in~\mmaInlineCell[]{Input}{list} to zero. 
\\\hline
\mmaInlineCell[]{Input}{\mmaDef{DefineParameters}[]}
&
Defines all SM parameters using as input scheme $\alpha_\mathrm{EM}=127.9^{-1}$, $G_F=1.16637 \times 10^{-5}\,\text{GeV}^{-1}$, and $M_Z=91.1876\,\text{GeV}$~\cite{Workman:2022ynf}. The widths of the SM bosons are taken as $\Gamma_Z=2.4952\,\text{GeV}$ and $\Gamma_W=2.085\,\text{GeV}$, and the Wolfenstein parameters are set to $\lambda=0.22650$, $A=0.790$, $\bar\rho=0.141$, and $\bar\eta=0.357$~\cite{Workman:2022ynf}. The parameters can be changed using the options \mmaInlineCell[]{Input}{"\(\alpha\)EM"}, \mmaInlineCell[]{Input}{"GF"}, \mmaInlineCell[]{Input}{"mZ"}, \mmaInlineCell[]{Input}{"\(\Gamma\)Z"}, \mmaInlineCell[]{Input}{"\(\Gamma\)W"}, and \mmaInlineCell[]{Input}{"Wolfenstein"}, where each option value should specify a single number apart from \mmaInlineCell[]{Input}{"Wolfenstein"} which must be a list of four numbers corresponding to $\lambda,A,\bar\rho,\bar\eta$. The CKM matrix elements are then given by
\[
    V_\mathrm{CKM} = 
    \begin{psmallmatrix}
        1-\lambda^2/2 & \lambda & A \lambda^3 (\rho-i\eta) \\
        -\lambda & 1-\lambda^2/2 & A \lambda^2 \\
        A \lambda^3 (1-\rho-i\eta) & -A \lambda^2 & 1
    \end{psmallmatrix},
\]
where $\rho=\bar{\rho}/(1-\lambda^2/2)$ and $\eta=\bar{\eta}/(1-\lambda^2/2)$.
To change for example the value of the electromagnetic fine structure constant one can use \mmaInlineCell[]{Input}{DefineParameters["\(\alpha\)EM"->1/137]}. Also the masses of BSM mediators, that have previously been defined with the \mmaInlineCell[]{Input}{InitializeModel} routine, can be modified using, e.g., \mmaInlineCell[]{Input}{DefineParameters[Mediators->\{"U1"->\{3000,10\}\}]} which would change the mass and with of a $U_1$~leptoquark to $3\,\mathrm{TeV}$ and $10\,\mathrm{GeV}$, respectively. To reset to the default values for all parameters use \mmaInlineCell[]{Input}{DefineParameters[Default]}. 
\\\hline
\mmaInlineCell[]{Input}{\mmaDef{GetParameters}[]}
&
Returns an \mmaInlineCell[]{Input}{\mmaDef{Association}} containing all SM parameters given as \mmaInlineCell[]{Input}{\mmaDef{Param}}, \mmaInlineCell[]{Input}{Mass}, \mmaInlineCell[]{Input}{Width}, or \mmaInlineCell[]{Input}{\mmaDef{Vckm}} and their currently used values. Also included are the values of all BSM mediator masses and widths.
\\\hline
\mmaInlineCell[]{Input}{DefineBasisAlignment[arg]}
&
Allows to specify the alignment of the mass basis to the flavor basis. To choose down alignment (default) \mmaInlineCell[]{Input}{arg="down"} has to be specified, whereas for an up-aligned basis \mmaInlineCell[]{Input}{arg="up"} must be specified. To work in any other basis one can give any (approximately) unitary ${3 \times 3}$~matrix as argument~\mmaInlineCell[]{Input}{arg}. In this case we define $V_d=$\,\mmaInlineCell[]{Input}{arg} and $V_u=V_d\cdot V_\mathrm{CKM}^\dagger$, where $V_{u,d}$ are the left-handed rotation matrices defined in eq.~\eqref{eq:qL_rotation_matrices}.
\\\hline
\mmaInlineCell[]{Input}{\mmaDef{SetPDF}["pdf_label"]}
&
Defines the PDF set that should be used for computing cross sections. The argument \mmaInlineCell[]{Input}{"pdf_label"} must correspond to one of the supported PDF sets. A~list of all included PDF sets can be obtained by \mmaInlineCell[]{Input}{\mmaDef{SetPDF}[]}. By default computations are performed using the average values of the \cmd{PDF4LHC15\_nnlo\_mc} PDF set~\cite{Butterworth:2015oua} (\mmaInlineCell[]{Input}{"pdf_label"="PDF4LHC15"}). For all supported PDF sets the average PDF is used as provided by \cmd{LHAPDF}~\cite{Buckley:2014ana}.
\\\hline
\mmaInlineCell[]{Input}{PythonExport["file",}

\quad \mmaInlineCell[]{Input}{\{f1,f2,...\},"search"]}

\mmaInlineCell[]{Input}{PythonExport["file",}

\quad \mmaInlineCell[]{Input}{\{f1,f2,...\}]}
&
Exports the \Mathematica expressions \mmaInlineCell[]{Input}{f1,f2,...} depending on Wilson coefficients or NP coupling constants to a python file named {\cmd{file.py}} using the Wilson coefficient exchange format \cmd{WCxf}~\cite{Aebischer:2017ugx}. For $d=8$ coefficients and coupling constants the conventions presented in appendix~\ref{app:WC-conventions} are used. The python file contains a variable \cmd{parameters} that contains all Wilson coefficients or coupling constants present in the given expressions. Also a function \cmd{file(C)} is included that depends on a dictionary \cmd{C} for the coefficients present and that returns the expressions \mmaInlineCell[]{Input}{f1,f2,...} as a list. The third argument \mmaInlineCell[]{Input}{"search"} is optional and must match, if given, the string specifying a specific search (see column two of table~\ref{tab:searches}). If given the created file contains a third function \cmd{search\_data()} which can be used to obtain information about binning, the observed number of events, the background estimates, and the background uncertainties, as provided by the experimental collaborations. By default the python file is stored in the directory given by \mmaInlineCell[]{Input}{NotebookDirectory[]}. An alternative location at \cmd{"path/to/dir"} can be specified through the option \mmaInlineCell[]{Input}{Directory->"path/to/dir"}.
\\\hline
\mmaInlineCell[]{Input}{HighPTLogo[]}
&
Returns a the \cmd{HighPT} logo as a \mmaInlineCell[]{Input}{Graphics} object that can be included in plots.
\\\hline
\mmaInlineCell[]{Input}{\mmaDef{TraditionalForm}}
&
Built in \Mathematica routine that can be used to display form factors, Wilson coefficients, coupling constants, etc. in a more readable format. Notice that displaying long expressions using \mmaInlineCell[]{Input}{\mmaDef{TraditionalForm}} is not recommended. 
\\\hline
\caption{Description of the most important routines used in \HighPT apart from the main routines discussed in sec.~\ref{sec:HighPT}.}
\label{tab:highpt}
\end{longtable}
\end{center}


\clearpage

\begin{center}
\begin{longtable}{| p{.35\textwidth} | p{.65\textwidth} |} 
\hline
\textbf{Variable} & \textbf{Description}
\\\hline
\mmaInlineCell[]{Input}{\mmaDef{WC}["label",\{\mmaUnd{\(\alpha\)},\mmaUnd{\(\beta\)},\mmaUnd{\(i\)},\mmaUnd{\(j\)}\}]}
\mmaInlineCell[]{Input}{\mmaDef{WC}["label",\{\mmaUnd{\(p\)},\mmaUnd{\(r\)}\}]}
&
Wilson coefficients of four- and two-fermion operators with lepton flavor indices \mmaInlineCell[]{Input}{\mmaUnd{\(\alpha\)},\mmaUnd{\(\beta\)}} and quark flavor indices \mmaInlineCell[]{Input}{\mmaUnd{\(i\)},\mmaUnd{\(j\)}}. The indices \mmaInlineCell[]{Input}{\mmaUnd{\(p\)},\mmaUnd{\(r\)}} can be either lepton or quark flavor indices depending on the operator. The type of the Wilson coefficient is given by \mmaInlineCell[]{Input}{"label"} which represents one of the Warsaw basis coefficients. A list of all Wilson coefficients is given in appendix~\ref{app:WC-conventions}.
For Hermitian four-fermion coefficients we automatically require \mmaInlineCell[]{Input}{\mmaUnd{\(\alpha\)}}$\leq$\mmaInlineCell[]{Input}{\mmaUnd{\(\beta\)}}, and if \mmaInlineCell[]{Input}{\mmaUnd{\(\alpha\)}}$=$\mmaInlineCell[]{Input}{\mmaUnd{\(\beta\)}} we further require \mmaInlineCell[]{Input}{\mmaUnd{\(i\)}}$\leq$\mmaInlineCell[]{Input}{\mmaUnd{\(j\)}}, to remove all redundancies in the couplings. For Hermitian two fermion coefficients we simply require \mmaInlineCell[]{Input}{\mmaUnd{\(p\)}}$\leq$\mmaInlineCell[]{Input}{\mmaUnd{\(r\)}}. Notice that this choice agrees with the \cmd{WCxf} convention~\cite{Aebischer:2017ugx} except for the coefficient $[\mathcal{C}_{qe}]_{\alpha\beta ij}$ in the Warsaw basis, which we dub $[\mathcal{C}_{eq}]_{\alpha\beta ij}$ to conveniently have the lepton flavor indices ordered before the quark flavor indices for all operators.
\\\hline
\mmaInlineCell[]{Input}{Coupling["label",\{\mmaUnd{\(p\)},\mmaUnd{\(r\)}\}]}
&
Denotes the coupling constants of a BSM mediator to fermions, where \mmaInlineCell[]{Input}{"label"} corresponds to the coupling label as given in tab.~\ref{tab:mediators} and \mmaInlineCell[]{Input}{\mmaUnd{\(p\)},\mmaUnd{\(r\)}} are the flavor indices.
\\\hline
\mmaInlineCell[]{Input}{FF[\mmaUnd{structure},\mmaUnd{type},\{\mmaUnd{\mmaSub{\(\chi\)}{\(\ell\)}},\mmaUnd{\mmaSub{\(\chi\)}{\(q\)}}\},}

\quad \mmaInlineCell[]{Input}{\{\mmaUnd{\mmaSub{\(\ell\)}{1}}[\mmaUnd{\(\alpha\)}],\mmaUnd{\mmaSub{\(\ell\)}{2}}[\mmaUnd{\(\beta\)}],\mmaUnd{\mmaSub{\(q\)}{1}}[\mmaUnd{\(i\)}],\mmaUnd{\mmaSub{\(q\)}{2}}[\mmaUnd{\(j\)}]\}]}
&
Form factor for the process ${\bar{q}_1^i q_2^j \to \ell_1^\alpha \bar{\ell}_2^\beta}$, 
where \mmaInlineCell[]{Input}{\mmaUnd{\mmaSub{\(q\)}{1,2}} \(\in\) \{\mmaDef{u,d}\}} and \mmaInlineCell[]{Input}{\mmaUnd{\mmaSub{\(\ell\)}{1,2}} \(\in\) \{\mmaDef{e,\(\nu\)}\}}. The chirality of the quark (lepton) current is given by \mmaInlineCell[]{Input}{\mmaUnd{\mmaSub{\(\chi\)}{\(q\)}}} (\mmaInlineCell[]{Input}{\mmaUnd{\mmaSub{\(\chi\)}{\(\ell\)}}}) with \mmaInlineCell[]{Input}{\mmaUnd{\mmaSub{\(\chi\)}{\(q,\ell\)}} \(\in\) \{Left,Right\}}
The Lorentz structure is given by \mmaInlineCell[]{Input}{\mmaUnd{structure} \(\in\) \{Scalar,Vector,Tensor,DipoleL,DipoleQ\}}. The argument \mmaInlineCell[]{Input}{\mmaUnd{type}} denotes whether the form factor is associated to a contact interaction or a mediator. In the former case, we have \mmaInlineCell[]{Input}{\mmaUnd{type}=\{"regular",\{\mmaUnd{\mmaSub{n}{s}},\mmaUnd{\mmaSub{n}{t}}\}\}} where the term scales with the partonic Mandelstam variables as~$\hat{s}$\mmaInlineCell[]{Input}{\mmaSup{\,}{\mmaSub{n}{s}}}$\hat{t}$\mmaInlineCell[]{Input}{\mmaSup{\,}{\mmaSub{n}{t}}}.
For the mediator form factors we have, e.g.,~\mmaInlineCell[]{Input}{\mmaUnd{type}=\{"ZBoson",SM\}} for the SM contribution by the $Z$~boson. The energy scaling of all form factors associated to mediators is always constant. However, we still separate the SM contributions from NP modifications which are denoted e.g. by \mmaInlineCell[]{Input}{\mmaUnd{type}=\{"ZBoson",0\}}.
\\\hline
\mmaInlineCell[]{Input}{\mmaDef{e}[\mmaUnd{\(\alpha\)}]}

\mmaInlineCell[]{Input}{\mmaDef{\(\nu\)}[\mmaUnd{\(\alpha\)}]}
& 
Charged leptons \mmaInlineCell[]{Input}{e} and neutrinos \mmaInlineCell[]{Input}{\(\nu\)} with flavor index~\mmaInlineCell[]{Input}{\mmaUnd{\(\alpha\)} \(\in\) \{1,2,3\}}. If \mmaInlineCell[]{Input}{\(\nu\)} is given without an index the summation over all flavors is implicitly understood for the computations of cross sections.
\\\hline
\mmaInlineCell[]{Input}{d[\mmaUnd{\(i\)}]}

\mmaInlineCell[]{Input}{u[\mmaUnd{\(i\)}]}
& 
Down-type \mmaInlineCell[]{Input}{d} and up-type \mmaInlineCell[]{Input}{u}\ quarks with flavor index~\mmaInlineCell[]{Input}{\mmaUnd{\(i\)} \(\in\) \{1,2,3\}}.
\\\hline
\mmaInlineCell[]{Input}{"Photon",\,"ZBoson",\,"WBoson"}
& 
Standard Model gauge bosons~$\gamma,Z,W^{\pm}$.
\\\hline
\mmaInlineCell[]{Input}{Scalar, Vector, Tensor,}
\mmaInlineCell[]{Input}{DipoleL, DipoleQ}
&
Arguments of form factors~(\mmaInlineCell[]{Input}{FF}) corresponding to the Lorentz structures of the associated operators. 
\\\hline
\mmaInlineCell[]{Input}{Left, Right}
&
Chiralities of the fermion bilinears associated to the form factors.
\\\hline
\mmaInlineCell[]{Input}{\mmaDef{Param}["param"]}
&
Denotes the parameter \mmaInlineCell[]{Input}{"param"}. The defined parameters are the electromagnetic fine structure constant (\mmaInlineCell[]{Input}{"param"="\(\alpha\)EM"}), Fermi's constant (\mmaInlineCell[]{Input}{"GF"}), the electroweak vacuum expectation value (\mmaInlineCell[]{Input}{"vev"}), and the sine (\mmaInlineCell[]{Input}{"sW"}) and cosine (\mmaInlineCell[]{Input}{"cW"}) of the weak mixing angle. The numerical value can be substituted using \mmaInlineCell[]{Input}{GetParameters[]}.
\\\hline
\mmaInlineCell[]{Input}{CKM}
&
CKM matrix with the element in the \mmaInlineCell[]{Input}{n}$^\text{th}$ row and \mmaInlineCell[]{Input}{m}$^\text{th}$ column given by \mmaInlineCell[]{Input}{Vckm[n,m]}.
\\\hline
\mmaInlineCell[]{Input}{Vckm[n,m]}
&
Denotes the element of the CKM matrix in the \mmaInlineCell[]{Input}{n}$^\text{th}$ row and \mmaInlineCell[]{Input}{m}$^\text{th}$ column. The numerical value can be substituted using \mmaInlineCell[]{Input}{GetParameters[]}.
\\\hline
\mmaInlineCell[]{Input}{\mmaDef{Mass}[\mmaUnd{\(\phi\)}]}
&
Mass of the mediator labeled by \mmaInlineCell[]{Input}{\mmaUnd{\(\phi\)}}, e.g. \mmaInlineCell[]{Input}{\mmaDef{Mass["ZBoson"]}} is the mass of the $Z$~bosons.
\\\hline
\mmaInlineCell[]{Input}{\mmaDef{Width}[\mmaUnd{\(\phi\)}]}
&
Width of the mediator labeled by \mmaInlineCell[]{Input}{\mmaUnd{\(\phi\)}}, e.g. \mmaInlineCell[]{Input}{\mmaDef{Width["ZBoson"]}} is the width of the $Z$~bosons.
\\\hline
\mmaInlineCell[]{Input}{\$ParallelHighPT}
&
Is a Boolean variable that specifies whether \mmaInlineCell[]{Input}{EventYield} and \mmaInlineCell[]{Input}{ChiSquareLHC} should be evaluated on all available Kernels in parallel (\mmaInlineCell[]{Input}{\$ParallelHighPT=True}) or just on a single Kernel (\mmaInlineCell[]{Input}{\$ParallelHighPT=False}). By default the multi-core mode is activated.\,\footnote{Notice that for parallelization to work, \Mathematica needs to be configured accordingly.}
\\\hline
\mmaInlineCell[]{Input}{\mmaDef{\$PrintingProcessInfo}}
&
Is a Boolean variable that specifies whether \mmaInlineCell[]{Input}{EventYield} and \mmaInlineCell[]{Input}{ChiSquareLHC} should print information about the specified search in the notebook when evaluated.
\\\hline
\caption{Description of all remaining variables in \HighPT. 
}
\label{tab:FF_WC}
\end{longtable}
\end{center}


\begin{center}
\begin{longtable}{| p{.35\textwidth} | p{.65\textwidth} |} 
\hline
\textbf{Option} & \textbf{Description}
\\\hline
\mmaInlineCell[]{Input}{\mmaDef{EFTorder} -> \mmaUnd{n}}
&
Specifies the order~$\mathcal{O}(\Lambda$\mmaInlineCell[]{Input}{\mmaSup{}{-n}}$)$ at which the EFT series in the NP scale~$\Lambda$ is truncated. The default value is defined by the last call of the \mmaInlineCell[]{Input}{\mmaDef{InitializeModel}} routine. Allowed values are \mmaInlineCell[]{Input}{\mmaUnd{n} \(\in\) \{0,2,4\}}. Used by: \mmaInlineCell[]{Input}{ChiSquareLHC}, \mmaInlineCell[]{Input}{EventYield}, \mmaInlineCell[]{Input}{CrossSection}, \mmaInlineCell[]{Input}{DifferentialCrossSection}, \mmaInlineCell[]{Input}{SubstituteFF}, \mmaInlineCell[]{Input}{InitializeModel}.
\\\hline
\mmaInlineCell[]{Input}{\mmaDef{OperatorDimension} -> \mmaUnd{d}}
&
Specifies the maximum mass dimension \mmaInlineCell[]{Input}{\mmaUnd{d}} of effective operators that are included in the SMEFT. The default value is defined by the last call of the \mmaInlineCell[]{Input}{\mmaDef{InitializeModel}} routine. Allowed values are \mmaInlineCell[]{Input}{\mmaUnd{d} \(\in\) \{4,6,8\}}. Used by: \mmaInlineCell[]{Input}{ChiSquareLHC}, \mmaInlineCell[]{Input}{EventYield}, \mmaInlineCell[]{Input}{CrossSection}, \mmaInlineCell[]{Input}{DifferentialCrossSection}, \mmaInlineCell[]{Input}{SubstituteFF}, \mmaInlineCell[]{Input}{InitializeModel}.
\\\hline
\mmaInlineCell[]{Input}{\mmaDef{EFTscale} -> \mmaUnd{\(\Lambda\)}}
&
Defines the EFT cutoff scale used when substituting in EFT Wilson coefficients for form factors. The default value is defined by the last call of the \mmaInlineCell[]{Input}{\mmaDef{InitializeModel}} routine. The EFT scale \mmaInlineCell[]{Input}{\mmaUnd{\(\Lambda\)}} must be given in units of~GeV. Used by:
\mmaInlineCell[]{Input}{ChiSquareLHC}, \mmaInlineCell[]{Input}{EventYield}, \mmaInlineCell[]{Input}{CrossSection}, \mmaInlineCell[]{Input}{DifferentialCrossSection}, \mmaInlineCell[]{Input}{SubstituteFF}, \mmaInlineCell[]{Input}{InitializeModel}.
\\\hline
\mmaInlineCell[]{Input}{\mmaDef{Coefficients} -> \mmaUnd{X}}
&
This option uses the \mmaInlineCell[]{Input}{SelectTerms} routine and allows to define which coefficients, i.e. form factors, Wilson coefficients, and/or coupling constants, should be kept in the result. By default (\mmaInlineCell[]{Input}{\mmaUnd{X}=\mmaDef{All}}) all terms are kept in the result. For example, to only retain the Wilson coefficients $[\cC_{lq}^{(1)}]_{3333}$ and $[\cC_{lq}^{(3)}]_{3333}$ in the result one can specify \mmaInlineCell[]{Input}{\mmaUnd{X}=\{\mmaDef{WC}["lq1",\{3,3,3,3\}],\mmaDef{WC}["lq3",\{3,3,3,3\}]\}}, all other occurrences of \mmaInlineCell[]{Input}{\mmaDef{WC}} are set to zero in this case. Notice that \mmaInlineCell[]{Input}{X} can also contain \mmaInlineCell[]{Input}{FF}, \mmaInlineCell[]{Input}{Coupling}, and patterns. Used by: \mmaInlineCell[]{Input}{ChiSquareLHC}, \mmaInlineCell[]{Input}{EventYield}, \mmaInlineCell[]{Input}{CrossSection}, \mmaInlineCell[]{Input}{DifferentialCrossSection}.
\\\hline
\mmaInlineCell[]{Input}{\mmaDef{FF} -> \mmaUnd{X}}
&
Defines whether the result should be returned in terms of form factors (\mmaInlineCell[]{Input}{X = True}), or 
whether form factors should be substituted in favor of Wilson coefficients or coupling constants (\mmaInlineCell[]{Input}{X = False}), where the latter is the default options.
Used by: \mmaInlineCell[]{Input}{ChiSquareLHC}, \mmaInlineCell[]{Input}{EventYield}, \mmaInlineCell[]{Input}{CrossSection}, \mmaInlineCell[]{Input}{DifferentialCrossSection}.
\\\hline
\mmaInlineCell[]{Input}{CombineBins -> X}
&
Allows to specify which bins of an experimental search should be combined before constructing the $\chi^2$~likelihood. Using this option can improve the implicit Gaussian approximation for constructing a $\chi^2$~likelihood if bins with no or very few events are present. The option value \mmaInlineCell[]{Input}{X} should be given as a list. For example using \mmaInlineCell[]{Input}{X = \{\{10,11\},\{12,13,14\}\}} combines the bin 10 with 11, as well as the bin 12 with 13 and~14. By default \mmaInlineCell[]{Input}{X = Default} experimental bins are combined in a way to ensure that at least 10 events are present in every resulting bin. Information about the default binning can be obtained by the \mmaInlineCell[]{Input}{LHCSearch} routine. Used by: \mmaInlineCell[]{Input}{ChiSquareLHC}.
\\\hline
\mmaInlineCell[]{Input}{Luminosity -> X}
&
Allows to change the luminosity for the computation of the event yield to allow for projections. The observed events and the background prediction are scaled by the ratio \mmaInlineCell[]{Input}{X}$/L_\mathrm{exp}$, where $L_\mathrm{exp}$ is the luminosity of the experimental search and~\mmaInlineCell[]{Input}{X} is the projected luminosity in units of~$\mathrm{fb}^{-1}$. If the \mmaInlineCell[]{Input}{Luminosity} option is specified, the number of expected events is substituted instead of the number of observed events for the projection. The scaling of the background uncertainty can be modified using the \mmaInlineCell[]{Input}{RescaleError} option.
Used by: \mmaInlineCell[]{Input}{ChiSquareLHC, EventYield}.
\\\hline
\mmaInlineCell[]{Input}{RescaleError -> X}
&
If the \mmaInlineCell[]{Input}{Luminosity} option is used to rescale the luminosity to obtain projections, this option allows to specify how background uncertainties are scaled. By default (\mmaInlineCell[]{Input}{X=True}) the background uncertainty~$\Delta\mathcal{N}^\mathrm{bkg}$ is rescaled by a factor $\smash{(L_\mathrm{proj}/L_\mathrm{exp})^{1/2}}$, where $L_\mathrm{proj}$ is the luminosity specified through the option~\mmaInlineCell[]{Input}{Luminosity} and $L_\mathrm{exp}$ is the experimental luminosity. If \mmaInlineCell[]{Input}{X=False} is specified the ratio $\Delta\mathcal{N}^\mathrm{bkg}/\mathcal{N}^\mathrm{bkg}$ is assumed to be constant, where $\mathcal{N}_B$ is the number of background events.
Used by: \mmaInlineCell[]{Input}{ChiSquareLHC}.
\\\hline
\mmaInlineCell[]{Input}{SM -> \mmaUnd{X}}
&
Specifies whether \mmaInlineCell[]{Input}{EventYield} considers the pure SM contribution to the expected number of events. If \mmaInlineCell[]{Input}{\mmaUnd{X}=False} (default) the SM contribution is omitted for \mmaInlineCell[]{Input}{\mmaUnd{X}=True} it is included. For the constructions of likelihoods the option \mmaInlineCell[]{Input}{SM->False} is used internally. Used by: \mmaInlineCell[]{Input}{EventYield}.
\\\hline
\mmaInlineCell[]{Input}{\mmaDef{MLLcuts} -> \{\mmaUnd{\mmaSubSup{m}{\(\ell\ell\)}{min}},\mmaUnd{\mmaSubSup{m}{\(\ell\ell\)}{max}}\}}
&
Defines the integration region \mmaInlineCell[]{Input}{\{\mmaUnd{\mmaSubSup{m}{\(\ell\ell\)}{min}},\mmaUnd{\mmaSubSup{m}{\(\ell\ell\)}{max}}\}} of the invariant mass of the dilepton system for the cross-section computation. Used by: \mmaInlineCell[]{Input}{CrossSection}.
\\\hline
\mmaInlineCell[]{Input}{\mmaDef{PTcuts} -> \{\mmaUnd{\mmaSubSup{p}{T}{min}},\mmaUnd{\mmaSubSup{p}{T}{max}}\}}
&
Defines the integration region \mmaInlineCell[]{Input}{\{\mmaUnd{\mmaSubSup{p}{T}{min}},\mmaUnd{\mmaSubSup{p}{T}{max}}\}} of the transverse momentum of the charged lepton for the cross-section computation. Used by: \mmaInlineCell[]{Input}{CrossSection}, \mmaInlineCell[]{Input}{DifferentialCrossSection}.
\\\hline
\caption{Common options used by \HighPT routines.}
\label{tab:highpt_options}
\end{longtable}
\end{center}

\section{On the statistical treatment of LHC data}
\label{app:stat}
In this appendix we compare the limits extracted with the $\chi^2$ test with those obtained using the CL$_s$ method \cite{Read:2000ru}.  The CL$_s$ method has the advantage of being a more robust statistic in regions of low sensitivity, like the high-$p_T$ tails of the distributions where the event number of signal and background events is low. This method is based on the {\it modified } $p$-value defined as CL$_{s}\equiv p_\theta/1-p_0$, where $p_\theta$ is the $p$-value for the signal at fixed parameter $\theta$ and $p_0$ is the $p$-value for the background-only hypothesis ($\theta=0$). For example, if the hypothesized signal satisfies CL$_s=0.05$, then it is considered excluded at $95$\% confidence level. At the LHC, the $p$-values are typically computed as
\begin{align}\label{eq:pval}
 p_\theta = \int_{q_{\theta}^{\rm obs}}^\infty\,\dd q_\theta\, f(q_\theta|\theta)
\end{align}
from the test statistic $q_\theta=-2\log \lambda(\theta)$, where $\lambda$ is the profiled Poissonian likelihood ratio \cite{1007.1727}, $q_{\theta}^{\rm obs}$ is the value of the observed test statistic and $f$ is the distribution of the test statistic for a signal at fixed $\theta$. For this particular case, the distribution $f$ in \eqref{eq:pval} can be conveniently approximated with asymptotic formulae, as shown in \cite{1007.1727}. 

\begin{figure}[t!]
 \centering
    \includegraphics[width=7.5cm]{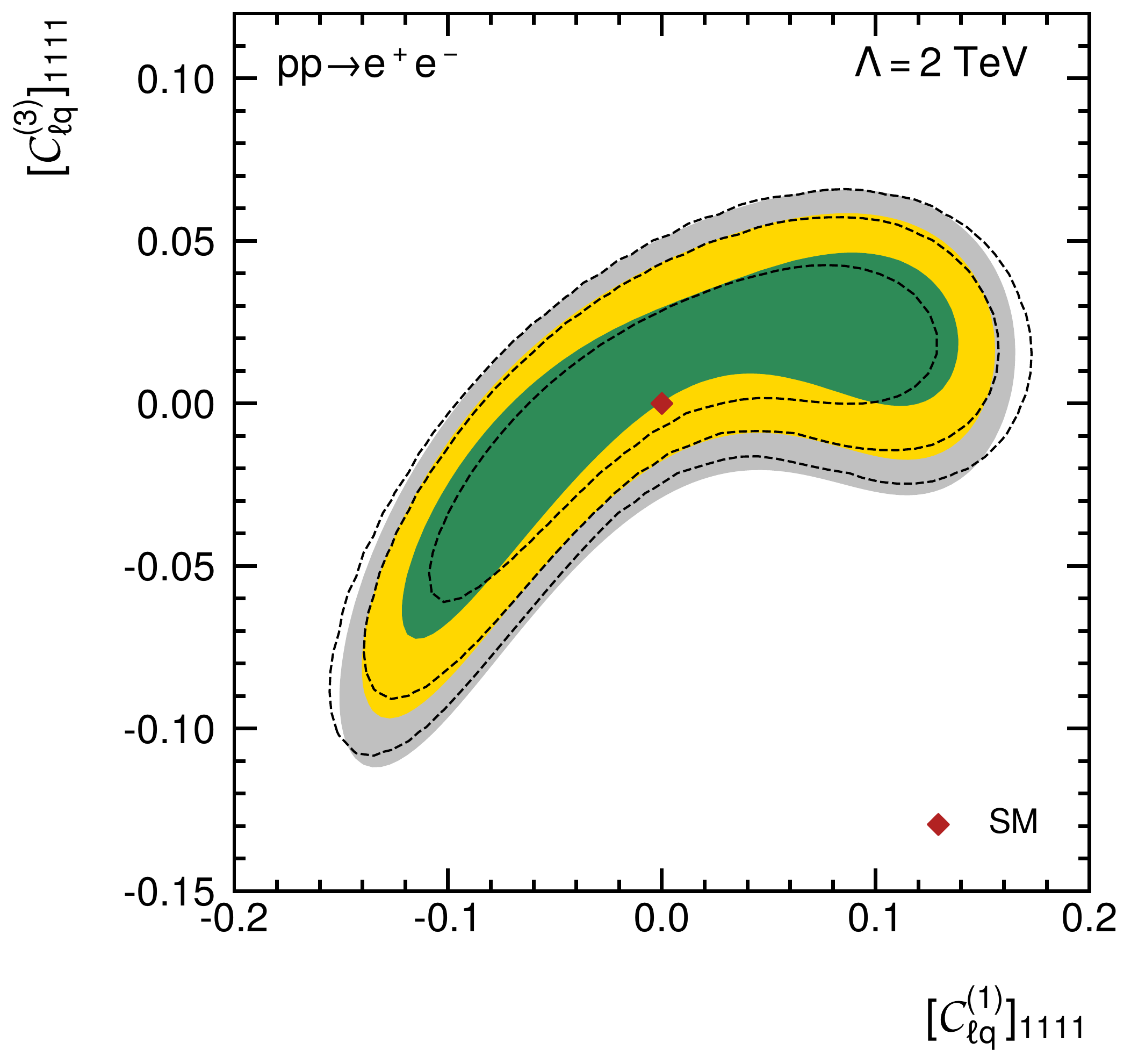}
    \hspace{0.25cm}
    \includegraphics[width=7.5cm]{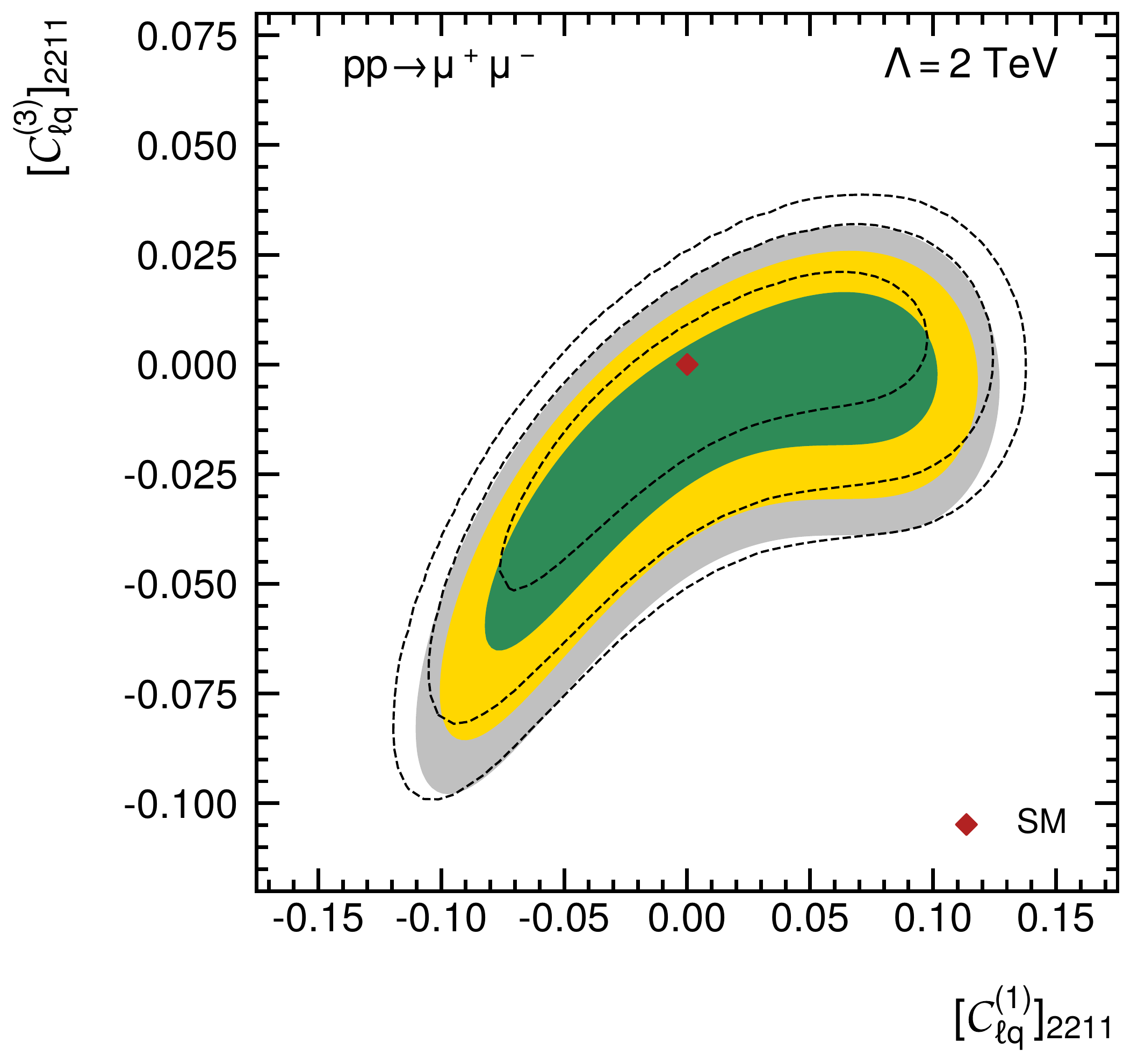} 
    \caption{Exclusion limits from $pp\to ee$ (left) and $pp\to \mu\mu$ (right) for Wilson coefficients of operators $\cO^{(1,3)}_{lq}$. The $1\sigma$, $2\sigma$ and $3\sigma$ regions for the $\chi^2$ statistic (green, yellow, grey) and the corresponding regions using the CL$_s$ method (dashed contours).}\label{fig:CLs_vs_chi2}
\end{figure}

In the following, we perform a comparison between the $\chi^2$ test and the CL$_s$ method by extracting exclusion limits for a simple New Physics model, using LHC dilepton data. For the benchmark signal, we use the SMEFT operators $\smash{\cO^{(1,3)}_{lq}}$ and turn on the Wilson coefficients $\smash{[\mathcal{C}^{(1,3)}_{lq}]_{\alpha\alpha 11}}$ for first generation valence quarks, muons and electrons ($\alpha=1,2$). For the CL$_s$, we compute the signal yields for $pp\to ee\,,\mu\mu$ at different values of the Wilson coefficients using {\tt HighPT}, and then use these as inputs in {\tt pyhf}~\cite{Heinrich:2021gyp} to extract the CL$_s$ exclusion limits based on the $\tilde q_\mu$ test statistic. The $1\sigma$, $2\sigma$ and $3\sigma$ regions for the $\chi^2$ (green, yellow, grey) and the CL$_s$ (dashed contours) for each dilepton mode are shown in fig.~\ref{fig:CLs_vs_chi2}. Overall, we find good agreement between the two statistical methods for these data sets.

\section{Conventions}
\label{app:WC-conventions}
To remove all the redundancies in the Wilson coefficients of Hermitian operators we requiring the following ordering of flavor indices: $\alpha \leq \beta$, and if $\alpha=\beta$ we furthermore demand $i \leq j$ for these type of operators. This selection corresponds to the choice by the \cmd{WCxf} format~\cite{Aebischer:2017ugx} for Wilson coefficients of dimension six, with one difference: the operator $[Q_{qe}]_{\alpha\beta ij}=(\bar{q}_i \gamma_\mu q_j)(e_\alpha \gamma^\mu e_\beta)$ in the Warsaw basis~\cite{Grzadkowski:2010es} is ordered by $i \leq j$ and if $i=j$ then also $\alpha \leq \beta$ in the \cmd{WCxf} conventions. We do note adopt this convention for this operator and therefore dub it $[Q_{eq}]_{\alpha\beta ij}$ instead. Notice the \mmaInlineCell[]{Input}{\mmaDef{PythonExport}} routine accounts for this miss match of conventions, and correctly translates the \HighPT results to the \cmd{WCxf} format. 
For the Wilson coefficient labels we also follow the \cmd{WCxf} conventions for $d=6$ operators, apart from the labeling of the Higgs, for which we use \cmd{H} rather than \cmd{phi}. A list of all $d=6$ coefficients is shown in tab.~\ref{tab:WC6}. For $d=8$ operators we define the labels used by \HighPT in tab.~\ref{tab:WC8}. We also use the same conventions for flavor indices of Hermitian operators as discussed above.

\clearpage

\begin{table}[h]
{\renewcommand{\arraystretch}{1.4}
\centering
\begin{tabular}{| p{1.7cm} | p{3.7cm} | p{4.4cm} |}
\hline
Coefficient & Operator & \HighPT label
\\\hline\hline
$[\mathcal{C}_{lq}^{(1)}]_{\alpha\beta ij}$ & $(\bar l_\alpha \gamma^\mu l_\beta)(\bar q_i \gamma_\mu q_j)$ & \mmaInlineCell[]{Input}{WC["lq1",\{\(\mmaUnd{\alpha},\mmaUnd{\beta},i,j\)\}]}
\\
$[\mathcal{C}_{lq}^{(3)}]_{\alpha\beta ij}$ & $(\bar l_\alpha \gamma^\mu\tau^I l_\beta)(\bar q_i \gamma_\mu\tau^I q_j)$ & \mmaInlineCell[]{Input}{WC["lq3",\{\(\mmaUnd{\alpha},\mmaUnd{\beta},i,j\)\}]}
\\
$[\mathcal{C}_{eu}]_{\alpha\beta ij}$ & $(\bar e_\alpha \gamma^\mu e_\beta)(\bar u_i \gamma_\mu u_j)$ & \mmaInlineCell[]{Input}{WC["eu",\{\(\mmaUnd{\alpha},\mmaUnd{\beta},i,j\)\}]}
\\
$[\mathcal{C}_{ed}]_{\alpha\beta ij}$ & $(\bar e_\alpha \gamma^\mu e_\beta)(\bar d_i \gamma_\mu d_j)$ & \mmaInlineCell[]{Input}{WC["ed",\{\(\mmaUnd{\alpha},\mmaUnd{\beta},i,j\)\}]}
\\
$[\mathcal{C}_{lu}]_{\alpha\beta ij}$ & $(\bar l_\alpha \gamma^\mu l_\beta)(\bar u_i \gamma_\mu u_j)$ & \mmaInlineCell[]{Input}{WC["lu",\{\(\mmaUnd{\alpha},\mmaUnd{\beta},i,j\)\}]}
\\
$[\mathcal{C}_{ld}]_{\alpha\beta ij}$ & $(\bar l_\alpha \gamma^\mu l_\beta)(\bar d_i \gamma_\mu d_j)$ & \mmaInlineCell[]{Input}{WC["ld",\{\(\mmaUnd{\alpha},\mmaUnd{\beta},i,j\)\}]}
\\
$[\mathcal{C}_{eq}]_{\alpha\beta ij}$ & $(\bar e_\alpha \gamma^\mu e_\beta)(\bar q_i \gamma_\mu q_j)$ & \mmaInlineCell[]{Input}{WC["eq",\{\(\mmaUnd{\alpha},\mmaUnd{\beta},i,j\)\}]}
\\
$[\mathcal{C}_{ledq}]_{\alpha\beta ij}$ & $(\bar l_\alpha e_\beta)(\bar d_i q_j)$ & \mmaInlineCell[]{Input}{WC["ledq",\{\(\mmaUnd{\alpha},\mmaUnd{\beta},i,j\)\}]}
\\
$[\mathcal{C}_{lequ}^{(1)}]_{\alpha\beta ij}$ & $(\bar l_\alpha e_\beta)\varepsilon(\bar q_i u_j)$ & \mmaInlineCell[]{Input}{WC["lequ1",\{\(\mmaUnd{\alpha},\mmaUnd{\beta},i,j\)\}]}
\\
$[\mathcal{C}_{lequ}^{(3)}]_{\alpha\beta ij}$ & $(\bar l_\alpha \sigma^{\mu\nu} e_\beta)\varepsilon(\bar q_i \sigma_{\mu\nu} u_j)$ & \mmaInlineCell[]{Input}{WC["lequ3",\{\(\mmaUnd{\alpha},\mmaUnd{\beta},i,j\)\}]}
\\\hline
\hline
$[\mathcal{C}_{eW}]_{\alpha\beta}$ & $(\bar l_\alpha \sigma^{\mu\nu} e_\beta)\, \tau^I H W^I_{\mu\nu}$ & \mmaInlineCell[]{Input}{WC["eW",\{\(\mmaUnd{\alpha},\mmaUnd{\beta}\)\}]}
\\
$[\mathcal{C}_{eB}]_{\alpha\beta}$ & $(\bar l_\alpha \sigma^{\mu\nu} e_\beta)\, H B_{\mu\nu}$ & \mmaInlineCell[]{Input}{WC["eB",\{\(\mmaUnd{\alpha},\mmaUnd{\beta}\)\}]}
\\
$[\mathcal{C}_{uW}]_{ij}$ & $(\bar q_i \sigma^{\mu\nu} u_j)\, \tau^I \widetilde{H} W^I_{\mu\nu}$ & \mmaInlineCell[]{Input}{WC["uW",\{\(i,j\)\}]}
\\
$[\mathcal{C}_{uB}]_{ij}$ & $(\bar q_i \sigma^{\mu\nu} u_j)\, \widetilde{H} B_{\mu\nu}$ & \mmaInlineCell[]{Input}{WC["uB",\{\(i,j\)\}]}
\\
$[\mathcal{C}_{dW}]_{ij}$ & $(\bar q_i \sigma^{\mu\nu} d_j)\, \tau^I H W^I_{\mu\nu}$ & \mmaInlineCell[]{Input}{WC["dW",\{\(i,j\)\}]}
\\
$[\mathcal{C}_{dB}]_{ij}$ & $(\bar q_i \sigma^{\mu\nu}\,d_j) H B_{\mu\nu}$ & \mmaInlineCell[]{Input}{WC["dB",\{\(i,j\)\}]}
\\\hline\hline
$[\mathcal{C}_{Hl}^{(1)}]_{\alpha\beta}$ & $(\bar l_\alpha \gamma^\mu l_\beta)(H^\dagger i\overleftrightarrow{D}_\mu H)$ & \mmaInlineCell[]{Input}{WC["Hl1",\{\(\mmaUnd{\alpha},\mmaUnd{\beta}\)\}]}
\\
$[\mathcal{C}_{Hl}^{(3)}]_{\alpha\beta}$ & $(\bar l_\alpha \gamma^\mu\tau^I l_\beta)(H^\dagger i\overleftrightarrow{D}^I_\mu H)$ & \mmaInlineCell[]{Input}{WC["Hl3",\{\(\mmaUnd{\alpha},\mmaUnd{\beta}\)\}]}
\\
$[\mathcal{C}_{He}]_{\alpha\beta}$ & $(\bar e_\alpha \gamma^\mu e_\beta)(H^\dagger i\overleftrightarrow{D}_\mu H)$ & \mmaInlineCell[]{Input}{WC["He",\{\(\mmaUnd{\alpha},\mmaUnd{\beta}\)\}]}
\\
$[\mathcal{C}_{Hq}^{(1)}]_{ij}$ & $(\bar q_i \gamma^\mu q_j)(H^\dagger i\overleftrightarrow{D}_\mu H)$ & \mmaInlineCell[]{Input}{WC["Hq1",\{\(i,j\)\}]}
\\
$[\mathcal{C}_{Hq}^{(3)}]_{ij}$ & $(\bar q_i \gamma^\mu\tau^I q_j)(H^\dagger i\overleftrightarrow{D}^I_\mu H)$ & \mmaInlineCell[]{Input}{WC["Hq3",\{\(i,j\)\}]}
\\
$[\mathcal{C}_{Hu}]_{ij}$ & $(\bar u_i \gamma^\mu u_j)(H^\dagger i\overleftrightarrow{D}_\mu H)$ & \mmaInlineCell[]{Input}{WC["Hu",\{\(i,j\)\}]}
\\
$[\mathcal{C}_{Hd}]_{ij}$ & $(\bar d_i \gamma^\mu d_j)(H^\dagger i\overleftrightarrow{D}_\mu H)$ & \mmaInlineCell[]{Input}{WC["Hd",\{\(i,j\)\}]}
\\
$[\mathcal{C}_{Hud}]_{ij}$ & $(\bar u_i \gamma^\mu d_j)(\widetilde H^\dagger iD_\nu H)$ & \mmaInlineCell[]{Input}{WC["Hud",\{\(i,j\)\}]}
\\\hline
\end{tabular}
\caption{List of all $d=6$ Wilson coefficients in the Warsaw basis~\cite{Grzadkowski:2010es} available in \HighPT. Lepton flavor indices are denoted $\alpha,\beta$, whereas quark flavor indices are labeled $i,j$. For the Hermitian $\psi^4$ operators we require the index ordering $\alpha \leq \beta$ and if $\alpha = \beta$ we furthermore request $i \leq j$. For Hermitian $\psi^2$~operators we require either $\alpha \leq \beta$ or $i \leq j$. The last column gives the labels used by \HighPT. The labels used by \cmd{PythonExport} are the corresponding \cmd{WCxf} labels.}
\label{tab:WC6}
}
\end{table}

\begin{center}
\begin{longtable}{| p{1.9cm} | p{6.5cm} | p{4.7cm} |}
\hline
Coefficient & Operator & \HighPT label
\\\hline\hline
$[\mathcal{C}_{l^2 q^2 D^2}^{(1)}]_{\alpha\beta ij}$ & $D^\nu (\bar l_\alpha \gamma^\mu l_\beta) D_\nu (\bar q_i \gamma_\mu q_j)$ & \mmaInlineCell[]{Input}{WC["l2q2D21",\{\(\mmaUnd{\alpha},\mmaUnd{\beta},i,j\)\}]}
\\
$[\mathcal{C}_{l^2 q^2 D^2}^{(2)}]_{\alpha\beta ij}$ & $(\bar l_\alpha \gamma^\mu \overleftrightarrow{D}^\nu l_\beta) (\bar q_i \gamma_\mu \overleftrightarrow{D}_\nu q_j)$ & \mmaInlineCell[]{Input}{WC["l2q2D22",\{\(\mmaUnd{\alpha},\mmaUnd{\beta},i,j\)\}]}
\\
$[\mathcal{C}_{l^2 q^2 D^2}^{(3)}]_{\alpha\beta ij}$ & $D^\nu (\bar l_\alpha \gamma^\mu \tau^I l_\beta) D_\nu (\bar q_i\gamma_\mu \tau^I q_j)$ & \mmaInlineCell[]{Input}{WC["l2q2D23",\{\(\mmaUnd{\alpha},\mmaUnd{\beta},i,j\)\}]}
\\
$[\mathcal{C}_{l^2 q^2 D^2}^{(4)}]_{\alpha\beta ij}$ & $(\bar l_\alpha \gamma^\mu \overleftrightarrow{D}^{I\nu} l_\beta) (\bar q_i \gamma_\mu \overleftrightarrow{D}^I_\nu q_j)$ & \mmaInlineCell[]{Input}{WC["l2q2D24",\{\(\mmaUnd{\alpha},\mmaUnd{\beta},i,j\)\}]}
\\
$[\mathcal{C}_{l^2 u^2 D^2}^{(1)}]_{\alpha\beta ij}$ & $D^\nu (\bar l_\alpha \gamma^\mu l_\beta) D_\nu (\bar u_i \gamma_\mu u_j)$ & \mmaInlineCell[]{Input}{WC["l2u2D21",\{\(\mmaUnd{\alpha},\mmaUnd{\beta},i,j\)\}]}
\\
$[\mathcal{C}_{l^2 u^2 D^2}^{(2)}]_{\alpha\beta ij}$ & $(\bar l_\alpha \gamma^\mu \overleftrightarrow{D}^\nu l_\beta) (\bar u_i \gamma_\mu \overleftrightarrow{D}_\nu u_j)$ & \mmaInlineCell[]{Input}{WC["l2u2D22",\{\(\mmaUnd{\alpha},\mmaUnd{\beta},i,j\)\}]}
\\
$[\mathcal{C}_{l^2 d^2 D^2}^{(1)}]_{\alpha\beta ij}$ & $D^\nu (\bar l_\alpha \gamma^\mu l_\beta) D_\nu (\bar d_i \gamma_\mu d_j)$ & \mmaInlineCell[]{Input}{WC["l2d2D21",\{\(\mmaUnd{\alpha},\mmaUnd{\beta},i,j\)\}]}
\\
$[\mathcal{C}_{l^2 d^2 D^2}^{(2)}]_{\alpha\beta ij}$ & $(\bar l_\alpha \gamma^\mu \overleftrightarrow{D}^\nu l_\beta) (\bar d_i \gamma_\mu \overleftrightarrow{D}_\nu d_j)$ & \mmaInlineCell[]{Input}{WC["l2d2D22",\{\(\mmaUnd{\alpha},\mmaUnd{\beta},i,j\)\}]}
\\
$[\mathcal{C}_{e^2 q^2 D^2}^{(1)}]_{\alpha\beta ij}$ & $D^\nu (\bar q_i \gamma^\mu q_j) D_\nu (\bar e_\alpha \gamma_\mu e_\beta)$ & \mmaInlineCell[]{Input}{WC["e2q2D21",\{\(\mmaUnd{\alpha},\mmaUnd{\beta},i,j\)\}]}
\\
$[\mathcal{C}_{e^2 q^2 D^2}^{(2)}]_{\alpha\beta ij}$ & $(\bar q_i \gamma^\mu \overleftrightarrow{D}^\nu q_j) (\bar e_\alpha \gamma_\mu \overleftrightarrow{D}_\nu e_\beta)$ & \mmaInlineCell[]{Input}{WC["e2q2D22",\{\(\mmaUnd{\alpha},\mmaUnd{\beta},i,j\)\}]}
\\
$[\mathcal{C}_{e^2 u^2 D^2}^{(1)}]_{\alpha\beta ij}$ & $D^\nu (\bar e_\alpha \gamma^\mu e_\beta) D_\nu (\bar u_i \gamma_\mu u_j)$ & \mmaInlineCell[]{Input}{WC["e2u2D21",\{\(\mmaUnd{\alpha},\mmaUnd{\beta},i,j\)\}]}
\\
$[\mathcal{C}_{e^2 u^2 D^2}^{(2)}]_{\alpha\beta ij}$ & $(\bar e_\alpha \gamma^\mu \overleftrightarrow{D}^\nu e_\beta) (\bar u_i \gamma_\mu \overleftrightarrow{D}_\nu u_j)$ & \mmaInlineCell[]{Input}{WC["e2u2D22",\{\(\mmaUnd{\alpha},\mmaUnd{\beta},i,j\)\}]}
\\
$[\mathcal{C}_{e^2 d^2 D^2}^{(1)}]_{\alpha\beta ij}$ & $D^\nu (\bar e_\alpha \gamma^\mu e_\beta) D_\nu (\bar d_i \gamma_\mu d_j)$ & \mmaInlineCell[]{Input}{WC["e2d2D21",\{\(\mmaUnd{\alpha},\mmaUnd{\beta},i,j\)\}]}
\\
$[\mathcal{C}_{e^2 d^2 D^2}^{(2)}]_{\alpha\beta ij}$ & $(\bar e_\alpha \gamma^\mu \overleftrightarrow{D}^\nu e_\beta) (\bar d_i \gamma_\mu \overleftrightarrow{D}_\nu d_j)$ & \mmaInlineCell[]{Input}{WC["e2d2D22",\{\(\mmaUnd{\alpha},\mmaUnd{\beta},i,j\)\}]}
\\\hline\hline
$[\mathcal{C}_{l^2 q^2 H^2}^{(1)}]_{\alpha\beta ij}$ & $(\bar l_\alpha \gamma^\mu l_\beta) (\bar q_i \gamma_\mu q_j) (H^\dag H)$ & \mmaInlineCell[]{Input}{WC["l2q2H21",\{\(\mmaUnd{\alpha},\mmaUnd{\beta},i,j\)\}]}
\\
$[\mathcal{C}_{l^2 q^2 H^2}^{(2)}]_{\alpha\beta ij}$ & $(\bar l_\alpha \gamma^\mu \tau^I l_\beta) (\bar q_i \gamma_\mu q_j) (H^\dag \tau^I H)$ & \mmaInlineCell[]{Input}{WC["l2q2H22",\{\(\mmaUnd{\alpha},\mmaUnd{\beta},i,j\)\}]}
\\
$[\mathcal{C}_{l^2 q^2 H^2}^{(3)}]_{\alpha\beta ij}$ & $(\bar l_\alpha \gamma^\mu \tau^I l_\beta) (\bar q_i \gamma_\mu \tau^I q_j) (H^\dag H)$ & \mmaInlineCell[]{Input}{WC["l2q2H23",\{\(\mmaUnd{\alpha},\mmaUnd{\beta},i,j\)\}]}
\\
$[\mathcal{C}_{l^2 q^2 H^2}^{(4)}]_{\alpha\beta ij}$ & $(\bar l_\alpha \gamma^\mu l_\beta) (\bar q_i \gamma_\mu \tau^I q_j) (H^\dag \tau^I H)$ & \mmaInlineCell[]{Input}{WC["l2q2H24",\{\(\mmaUnd{\alpha},\mmaUnd{\beta},i,j\)\}]}
\\
$[\mathcal{C}_{l^2 q^2 H^2}^{(5)}]_{\alpha\beta ij}$ & $\epsilon^{I\!J\!K} (\bar l_\alpha \gamma^\mu \tau^I l_\beta) (\bar q_i \gamma_\mu \tau^J q_j) (H^\dag \tau^K H)$ & \mmaInlineCell[]{Input}{WC["l2q2H25",\{\(\mmaUnd{\alpha},\mmaUnd{\beta},i,j\)\}]}
\\
$[\mathcal{C}_{l^2 u^2 H^2}^{(1)}]_{\alpha\beta ij}$ & $(\bar l_\alpha \gamma^\mu l_\beta) (\bar u_i \gamma_\mu u_j) (H^\dag H)$ & \mmaInlineCell[]{Input}{WC["l2u2H21",\{\(\mmaUnd{\alpha},\mmaUnd{\beta},i,j\)\}]}
\\
$[\mathcal{C}_{l^2 u^2 H^2}^{(2)}]_{\alpha\beta ij}$ & $(\bar l_\alpha \gamma^\mu \tau^I l_\beta) (\bar u_i \gamma_\mu u_j) (H^\dag \tau^I H)$ & \mmaInlineCell[]{Input}{WC["l2u2H22",\{\(\mmaUnd{\alpha},\mmaUnd{\beta},i,j\)\}]}
\\
$[\mathcal{C}_{l^2 d^2 H^2}^{(1)}]_{\alpha\beta ij}$ & $(\bar l_\alpha \gamma^\mu l_\beta) (\bar d_i \gamma_\mu d_j) (H^\dag H)$ & \mmaInlineCell[]{Input}{WC["l2d2H21",\{\(\mmaUnd{\alpha},\mmaUnd{\beta},i,j\)\}]}
\\
$[\mathcal{C}_{l^2 d^2 H^2}^{(2)}]_{\alpha\beta ij}$ & $(\bar l_\alpha \gamma^\mu \tau^I l_\beta) (\bar d_i \gamma_\mu d_j) (H^\dag \tau^I H)$ & \mmaInlineCell[]{Input}{WC["l2d2H22",\{\(\mmaUnd{\alpha},\mmaUnd{\beta},i,j\)\}]}
\\
$[\mathcal{C}_{e^2 q^2 H^2}^{(1)}]_{\alpha\beta ij}$ & $(\bar q_i \gamma^\mu q_j) (\bar e_\alpha \gamma_\mu e_\beta) (H^\dag H)$ & \mmaInlineCell[]{Input}{WC["e2q2H21",\{\(\mmaUnd{\alpha},\mmaUnd{\beta},i,j\)\}]}
\\
$[\mathcal{C}_{e^2 q^2 H^2}^{(2)}]_{\alpha\beta ij}$ & $(\bar q_i \gamma^\mu \tau^I q_j) (\bar e_\alpha \gamma_\mu e_\beta) (H^\dag \tau^I H)$ & \mmaInlineCell[]{Input}{WC["e2q2H22",\{\(\mmaUnd{\alpha},\mmaUnd{\beta},i,j\)\}]}
\\
$[\mathcal{C}_{e^2 u^2 H^2}]_{\alpha\beta ij}$ & $(\bar e_\alpha \gamma^\mu e_\beta) (\bar u_i \gamma_\mu u_j) (H^\dag H)$ & \mmaInlineCell[]{Input}{WC["e2u2H2",\{\(\mmaUnd{\alpha},\mmaUnd{\beta},i,j\)\}]}
\\
$[\mathcal{C}_{e^2 d^2 H^2}]_{\alpha\beta ij}$ & $(\bar e_\alpha \gamma^\mu e_\beta) (\bar d_i \gamma_\mu d_j) (H^\dag H)$ & \mmaInlineCell[]{Input}{WC["e2d2H2",\{\(\mmaUnd{\alpha},\mmaUnd{\beta},i,j\)\}]}
\\\hline
\hline
$[\mathcal{C}_{l^2 H^2 D^3}^{(1)}]_{\alpha\beta}$ & $ i (\bar l_\alpha \gamma^{\mu} D^\nu l_\beta) \,(D_{(\mu}D_{\nu)} H)^{\dag} H$ & \mmaInlineCell[]{Input}{WC["l2H2D31",\{\(\mmaUnd{\alpha},\mmaUnd{\beta}\)\}]}
\\
$[\mathcal{C}_{l^2 H^2 D^3}^{(2}]_{\alpha\beta}$ & $ i (\bar l_\alpha \gamma^{\mu} D^\nu l_\beta) \,H^{\dag} (D_{(\mu}D_{\nu)}  H)$ & \mmaInlineCell[]{Input}{WC["l2H2D32",\{\(\mmaUnd{\alpha},\mmaUnd{\beta}\)\}]}
\\
$[\mathcal{C}_{l^2 H^2 D^3}^{(3)}]_{\alpha\beta}$ & $ i (\bar l_\alpha \gamma^{\mu}\tau^I D^\nu l_\beta)\, (D_{(\mu}D_{\nu)} H)^{\dag}\tau^I H)$ & \mmaInlineCell[]{Input}{WC["l2H2D33",\{\(\mmaUnd{\alpha},\mmaUnd{\beta}\)\}]}
\\
$[\mathcal{C}_{l^2 H^2 D^3}^{(4)}]_{\alpha\beta}$ & $ i (\bar l_\alpha \gamma^{\mu}\tau^I D^\nu l_\beta) \, H^{\dag} \tau^I (D_{(\mu}D_{\nu)} H)$ & \mmaInlineCell[]{Input}{WC["l2H2D34",\{\(\mmaUnd{\alpha},\mmaUnd{\beta}\)\}]}
\\
$[\mathcal{C}_{e^2 H^2 D^3}^{(1)}]_{\alpha\beta}$ & $ i (\bar e_\alpha \gamma^{\mu} D^\nu e_\beta) \,(D_{(\mu}D_{\nu)} H)^{\dag} H)$ & \mmaInlineCell[]{Input}{WC["e2H2D31",\{\(\mmaUnd{\alpha},\mmaUnd{\beta}\)\}]}
\\
$[\mathcal{C}_{e^2 H^2 D^3}^{(2)}]_{\alpha\beta}$ & $ i (\bar e_\alpha \gamma^{\mu} D^\nu e_\beta) \,H^{\dag} (D_{(\mu}D_{\nu)}  H)$ & \mmaInlineCell[]{Input}{WC["e2H2D32",\{\(\mmaUnd{\alpha},\mmaUnd{\beta}\)\}]}
\\
$[\mathcal{C}_{q^2 H^2 D^3}^{(1)}]_{ij}$ & $ i (\bar q_i \gamma^{\mu} D^\nu q_j) \,(D_{(\mu}D_{\nu)} H)^{\dag} H$ & \mmaInlineCell[]{Input}{WC["q2H2D31",\{\(i,j\)\}]}
\\
$[\mathcal{C}_{q^2 H^2 D^3}^{(2)}]_{ij}$ & $ i (\bar q_i \gamma^{\mu} D^\nu q_j) \,H^{\dag} (D_{(\mu}D_{\nu)}  H)$ & \mmaInlineCell[]{Input}{WC["q2H2D32",\{\(i,j\)\}]}
\\
$[\mathcal{C}_{q^2 H^2 D^3}^{(3)}]_{ij}$ & $ i (\bar q_i \gamma^{\mu}\tau^I D^\nu q_j)\, (D_{(\mu}D_{\nu)} H)^{\dag}\tau^I H$ & \mmaInlineCell[]{Input}{WC["q2H2D33",\{\(i,j\)\}]}
\\
$[\mathcal{C}_{q^2 H^2 D^3}^{(4)}]_{ij}$ & $ i (\bar q_i \gamma^{\mu}\tau^I D^\nu q_j) \, H^{\dag} \tau^I (D_{(\mu}D_{\nu)} H)$ & \mmaInlineCell[]{Input}{WC["q2H2D34",\{\(i,j\)\}]}
\\
$[\mathcal{C}_{u^2 H^2 D^3}^{(1)}]_{ij}$ & $ i (\bar u_i \gamma^{\mu} D^\nu u_j) \, (D_{(\mu}D_{\nu)} H)^{\dag} H$ & \mmaInlineCell[]{Input}{WC["u2H2D31",\{\(i,j\)\}]}
\\
$[\mathcal{C}_{u^2 H^2 D^3}^{(2)}]_{ij}$ & $ i (\bar u_i \gamma^{\mu} D^\nu u_j) \,H^{\dag} (D_{(\mu}D_{\nu)}  H)$ & \mmaInlineCell[]{Input}{WC["u2H2D32",\{\(i,j\)\}]}
\\
$[\mathcal{C}_{d^2 H^2 D^3}^{(1)}]_{ij}$ & $ i (\bar d_i \gamma^{\mu} D^\nu d_j) \,(D_{(\mu}D_{\nu)} H)^{\dag} H$ & \mmaInlineCell[]{Input}{WC["d2H2D31",\{\(i,j\)\}]}
\\
$[\mathcal{C}_{d^2 H^2 D^3}^{(2)}]_{ij}$ & $ i (\bar d_i \gamma^{\mu} D^\nu d_j) \,H^{\dag} (D_{(\mu}D_{\nu)}  H)$ & \mmaInlineCell[]{Input}{WC["d2H2D32",\{\(i,j\)\}]}
\\\hline\hline
$[\mathcal{C}_{l^2 H^4 D}^{(1)}]_{\alpha\beta}$ & $ i (\bar l_\alpha \gamma^{\mu} l_\beta) (H^{\dag} \overleftrightarrow{D}_{\mu} H) (H^{\dag} H)$ & \mmaInlineCell[]{Input}{WC["l2H4D1",\{\(\mmaUnd{\alpha},\mmaUnd{\beta}\)\}]}
\\
$[\mathcal{C}_{l^2 H^4 D}^{(2)}]_{\alpha\beta}$ & $ i (\bar l_\alpha \gamma^{\mu} \tau^I l_\beta) [(H^{\dag} \overleftrightarrow{D}_{\mu}^I H) (H^{\dag} H) + (H^{\dag} \overleftrightarrow{D}_{\mu} H) (H^{\dag} \tau^I H)]$ & \mmaInlineCell[]{Input}{WC["l2H4D2",\{\(\mmaUnd{\alpha},\mmaUnd{\beta}\)\}]}
\\
$[\mathcal{C}_{l^2 H^4 D}^{(3)}]_{\alpha\beta}$ & $ \epsilon^{IJK} (\bar l_\alpha \gamma^{\mu} \tau^I l_\beta) (H^{\dag} \overleftrightarrow{D}_{\mu}^J H) (H^{\dag} \tau^K H)$ & \mmaInlineCell[]{Input}{WC["l2H4D3",\{\(\mmaUnd{\alpha},\mmaUnd{\beta}\)\}]}
\\
$[\mathcal{C}_{l^2 H^4 D}^{(4)}]_{\alpha\beta}$ & $ \epsilon^{IJK} (\bar l_\alpha \gamma^{\mu} \tau^I l_\beta) (H^{\dag} \tau^J H) (D_{\mu} H)^{\dag} \tau^K H$ & \mmaInlineCell[]{Input}{WC["l2H4D4",\{\(\mmaUnd{\alpha},\mmaUnd{\beta}\)\}]}
\\
$[\mathcal{C}_{q^2 H^4 D}^{(1)}]_{ij}$ & $ i (\bar q_i \gamma^{\mu} q_j) (H^{\dag} \overleftrightarrow{D}_{\mu} H) (H^{\dag} H)$ & \mmaInlineCell[]{Input}{WC["q2H4D1",\{\(i,j\)\}]}
\\
$[\mathcal{C}_{q^2 H^4 D}^{(2)}]_{ij}$ & $ i (\bar q_i \gamma^{\mu} \tau^I q_j) [(H^{\dag} \overleftrightarrow{D}_{\mu}^I H) (H^{\dag} H) + (H^{\dag} \overleftrightarrow{D}_{\mu} H) (H^{\dag} \tau^I H)]$ & \mmaInlineCell[]{Input}{WC["q2H4D2",\{\(i,j\)\}]}
\\
$[\mathcal{C}_{q^2 H^4 D}^{(3)}]_{ij}$ & $ i \epsilon^{IJK} (\bar q_i \gamma^{\mu} \tau^I q_j) (H^{\dag} \overleftrightarrow{D}_{\mu}^J H) (H^{\dag} \tau^K H)$ & \mmaInlineCell[]{Input}{WC["q2H4D3",\{\(i,j\)\}]}
\\
$[\mathcal{C}_{q^2 H^4 D}^{(4)}]_{ij}$ & $ \epsilon^{IJK} (\bar q_i \gamma^{\mu} \tau^I q_j) (H^{\dag} \tau^J H) (D_{\mu} H)^{\dag} \tau^K H$ & \mmaInlineCell[]{Input}{WC["q2H4D4",\{\(i,j\)\}]}
\\
$[\mathcal{C}_{e^2 H^4 D}]^{\alpha\beta}$ & $ i (\bar e_\alpha \gamma^{\mu} e_\beta) (H^{\dag} \overleftrightarrow{D}_{\mu} H) (H^{\dag} H)$ & \mmaInlineCell[]{Input}{WC["e2H4D",\{\(\mmaUnd{\alpha},\mmaUnd{\beta}\)\}]}
\\
$[\mathcal{C}_{u^2 H^4 D}]_{ij}$ & $ i (\bar u_i \gamma^{\mu} u_j) (H^{\dag} \overleftrightarrow{D}_{\mu} H) (H^{\dag} H)$ & \mmaInlineCell[]{Input}{WC["u2H4D",\{\(i,j\)\}]}
\\
$[\mathcal{C}_{d^2 H^4 D}]_{ij}$ & $ i (\bar d_i \gamma^{\mu} d_j) (H^{\dag} \overleftrightarrow{D}_{\mu} H) (H^{\dag} H)$ & \mmaInlineCell[]{Input}{WC["d2H4D",\{\(i,j\)\}]}
\\\hline
\caption{List of all $d=8$ Wilson coefficients in the basis of ref.~\cite{Murphy:2020rsh} available in \HighPT. Lepton flavor indices are denoted $\alpha,\beta$, whereas quark flavor indices are labeled $i,j$. 
Notice that the coefficients $\smash{\cC_{q^2 e^2 D^2}^{(1,2)}}$ in the basis of ref.~\cite{Murphy:2020rsh} have been relabeled to $\smash{\cC_{e^2 q^2 D^2}^{(1,2)}}$ in our convention to have lepton flavor indices before quark flavor indices for all operators. For the same reason we have also relabeled $\smash{\cC_{q^2 e^2 H^2}^{(1,2)}}$ to $\smash{\cC_{e^2 q^2 H^2}^{(1,2)}}$.
For the Hermitian $\psi^4$ operators we require the index ordering $\alpha \leq \beta$ and if $\alpha = \beta$ we furthermore request $i \leq j$. For Hermitian $\psi^2$~operators we require either $\alpha \leq \beta$ or $i \leq j$. The last column gives the labels used by \HighPT. The labels used by \cmd{PythonExport} are chosen analogous to the \cmd{WCxf} labels for $d=6$ operators.}
\label{tab:WC8}
\end{longtable}
\end{center}

\begin{table}[h]
{\renewcommand{\arraystretch}{1.2}
\centering
\begin{tabular}{| p{1.6cm} | p{4.5cm} | p{2.1cm} |}
\hline
Coupling & \HighPT label & \cmd{python} label
\\\hline\hline
$[y_1^L]_{i\alpha}$ & \mmaInlineCell[]{Input}{Coupling["y1L",\{i,\mmaUnd{\(\alpha\)}\}]} & \cmd{y1L\_i$\alpha$}
\\
$[y_1^R]_{i\alpha}$ & \mmaInlineCell[]{Input}{Coupling["y1R",\{i,\mmaUnd{\(\alpha\)}\}]} & \cmd{y1R\_i$\alpha$}
\\
$[\tilde{y}_1^R]_{i\alpha}$ & \mmaInlineCell[]{Input}{Coupling["y1Rt",\{i,\mmaUnd{\(\alpha\)}\}]} & \cmd{y1Rt\_i$\alpha$}
\\
$[x_1^L]_{i\alpha}$ & \mmaInlineCell[]{Input}{Coupling["x1L",\{i,\mmaUnd{\(\alpha\)}\}]} & \cmd{x1L\_i$\alpha$}
\\
$[x_1^R]_{i\alpha}$ & \mmaInlineCell[]{Input}{Coupling["x1R",\{i,\mmaUnd{\(\alpha\)}\}]} & \cmd{x1R\_i$\alpha$}
\\
$[\tilde{x}_1^R]_{i\alpha}$ & \mmaInlineCell[]{Input}{Coupling["x1Rt",\{i,\mmaUnd{\(\alpha\)}\}]} & \cmd{x1Rt\_i$\alpha$}
\\
$[y_2^L]_{i\alpha}$ & \mmaInlineCell[]{Input}{Coupling["y2L",\{i,\mmaUnd{\(\alpha\)}\}]} & \cmd{y2L\_i$\alpha$}
\\
$[y_2^R]_{i\alpha}$ & \mmaInlineCell[]{Input}{Coupling["y2R",\{i,\mmaUnd{\(\alpha\)}\}]} & \cmd{y2R\_i$\alpha$}
\\
$[\tilde{y}_2^L]_{i\alpha}$ & \mmaInlineCell[]{Input}{Coupling["y2Lt",\{i,\mmaUnd{\(\alpha\)}\}]} & \cmd{y2Lt\_i$\alpha$}
\\
$[x_2^L]_{i\alpha}$ & \mmaInlineCell[]{Input}{Coupling["x2L",\{i,\mmaUnd{\(\alpha\)}\}]} & \cmd{x2L\_i$\alpha$}
\\
$[x_2^R]_{i\alpha}$ & \mmaInlineCell[]{Input}{Coupling["x2R",\{i,\mmaUnd{\(\alpha\)}\}]} & \cmd{x2R\_i$\alpha$}
\\
$[\tilde{x}_2^L]_{i\alpha}$ & \mmaInlineCell[]{Input}{Coupling["x2Lt",\{i,\mmaUnd{\(\alpha\)}\}]} & \cmd{x2Lt\_i$\alpha$}
\\
$[y_3^L]_{i\alpha}$ & \mmaInlineCell[]{Input}{Coupling["y3L",\{i,\mmaUnd{\(\alpha\)}\}]} & \cmd{y3L\_i$\alpha$}
\\
$[x_3^L]_{i\alpha}$ & \mmaInlineCell[]{Input}{Coupling["x3L",\{i,\mmaUnd{\(\alpha\)}\}]} & \cmd{x3L\_i$\alpha$}
\\\hline
\end{tabular}
\caption{Couplings of NP mediators as defined in the Lagrangians of tab.~\ref{tab:mediators}. Quark flavor indices are denoted with~$i$, whereas lepton flavor indices are labeled~$\alpha$. The second column denotes the labels used by \HighPT, whereas the last column lists the labels used by the \cmd{PythonExport} routine.}
\label{tab:couplings}
}
\end{table}

\clearpage
{
\footnotesize
\bibliography{main.bib}
}

\end{document}